\documentclass[aps,prd,superscriptaddress,amsmath,amssymb,eqsecnum,nofootinbib,preprintnumbers,nolongbibliography,10pt]{revtex4-2}
\usepackage{url}
\usepackage{silence}
\usepackage[colorlinks=true,
    linkcolor=blue,
    citecolor=blue,
    urlcolor=blue]{hyperref}
 \usepackage{amsmath}
\usepackage{graphicx}
\usepackage{epstopdf}
\usepackage{color}
\usepackage[T1]{fontenc}
\usepackage[utf8]{inputenc}
\usepackage[toc,page]{appendix}
\usepackage[dvipsnames]{xcolor}
\usepackage[normalem]{ulem}

\newcommand{\bfeq}[1]

\makeatletter
\providecommand*{\diff}%
{\@ifnextchar^{\DIfF}{\DIfF^{}}}
\def\DIfF^#1{%
\mathop{\mathrm{\mathstrut d}}%
\nolimits^{#1}\gobblespace}
\def\gobblespace{%
\futurelet\diffarg\opspace}
\def\opspace{%
\let\DiffSpace\!%
\ifx\diffarg(%
\let\DiffSpace\relax
\else
\ifx\diffarg[%
\let\DiffSpace\relax
\else
\ifx\diffarg\{%
\let\DiffSpace\relax
\fi\fi\fi\DiffSpace}

\usepackage{xcolor}
\usepackage{scalerel}

\usepackage{tikz}
\usetikzlibrary{svg.path}
\definecolor{darkmagenta}{HTML}{8B008B}
\definecolor{orcidlogocol}{HTML}{A6CE39}
\tikzset{
  orcidlogo/.pic={
    \fill[orcidlogocol] svg{M256,128c0,70.7-57.3,128-128,128C57.3,256,0,198.7,0,128C0,57.3,57.3,0,128,0C198.7,0,256,57.3,256,128z};
    \fill[white] svg{M86.3,186.2H70.9V79.1h15.4v48.4V186.2z}
                 svg{M108.9,79.1h41.6c39.6,0,57,28.3,57,53.6c0,27.5-21.5,53.6-56.8,53.6h-41.8V79.1z M124.3,172.4h24.5c34.9,0,42.9-26.5,42.9-39.7c0-21.5-13.7-39.7-43.7-39.7h-23.7V172.4z}
                 svg{M88.7,56.8c0,5.5-4.5,10.1-10.1,10.1c-5.6,0-10.1-4.6-10.1-10.1c0-5.6,4.5-10.1,10.1-10.1C84.2,46.7,88.7,51.3,88.7,56.8z};
  }
}

\newcommand\orcidicon[1]{\href{https://orcid.org/#1}{\mbox{\scalerel*{
\begin{tikzpicture}[yscale=-1, scale=0.05, transform shape]
\pic{orcidlogo};
\end{tikzpicture}
}{|}}}}
\begin{document}


\title{Rotation in Einstein--ModMax theory}

\author{Jos\'e Barrientos\,$^{\orcidicon{0000-0003-3445-8151}}$}
\email{barrientos@math.cas.cz}
\affiliation{Institute of Mathematics of the Czech Academy of Sciences, {\v Z}itn{\'a} 25, 115 67 Praha 1, Czech Republic}
\affiliation{Vicerrector\'ia de Investigaci\'on y Postgrado, Universidad de La Serena, La Serena 1700000, Chile}

\author{Adolfo Cisterna\,$^{\orcidicon{0000-0002-6949-7882}}$}
\email{adolfo.cisterna@mff.cuni.cz}
\affiliation{Sede Esmeralda, Universidad de Tarapac{\'a}, Avenida Luis Emilio Recabarren 2477, Iquique, Chile}
\affiliation{Institute of Theoretical Physics, Faculty of Mathematics and Physics, Charles University, V Hole{\v s}ovi{\v c}k{\' a}ch 2, 180 00 Praha 8, Czech Republic}

\author{Carlos Herdeiro\,$^{\orcidicon{0000-0002-9619-2013}}$}
\email{herdeiro@ua.pt}
\affiliation{Departamento de Matem\'atica da Universidade de Aveiro and Centre for Research and Development  in Mathematics and Applications (CIDMA), Campus de Santiago, 3810-193 Aveiro, Portugal}

\begin{abstract}
We obtain analytic perturbative solutions describing nonextremal rotating Einstein--ModMax configurations in four dimensions. The construction retains independent frame-dragging and magnetic profiles, accommodating the nonlinear constitutive response that obstructs a simple charge-rescaled Kerr--Newman solution. At first order in the rotation parameter $a$, we derive closed expressions in mixed orders $\mathcal O(a\gamma)$ and $\mathcal O(a\gamma^2)$, where $\gamma$ is the ModMax coupling. {The resulting corrections involve logarithms, dilogarithms and multiple polylogarithms.} Electromagnetic duality generates the corresponding dyonic solutions. We also extend the electric sector through $\mathcal O(a^2\gamma)$, determining the monopole and quadrupole corrections to the geometry and electric potential in explicit analytic radial quadratures. {For fixed physical parameters, horizon regularity and asymptotic normalisation determine the integration constants, and the coupled field equations are satisfied at the retained perturbative orders.} Beyond the small-coupling expansion, we construct an exact representation of the first-order rotational solution for arbitrary finite $\gamma$ using convergent exterior Frobenius series, with recursive coefficients and a connection constant fixed by horizon regularity. {The rotation-induced magnetic field exhibits an anomalous fractional-power asymptotic decay, which we trace to the anisotropic constitutive response of ModMax around an electric Coulomb background. The same anomalous dipole behaviour arises already in flat spacetime, providing a physical explanation for the failure of Kerr--Newman-type electromagnetic ans\"atze. The expansion of the fractional power in $\gamma$ accounts for the logarithmic terms in the perturbative solutions, while the geometry remains asymptotically flat.}
\end{abstract}
\maketitle

\section{Introduction}

Nonlinear electrodynamics extends Maxwell theory by replacing its linear constitutive law with electromagnetic self-interactions compatible with gauge and Lorentz invariance. Its original motivation was to resolve the divergent electrostatic self-energy of a point charge, as achieved in Born--Infeld theory~\cite{Born:1934gh}. Nonlinear interactions also arise in effective descriptions of fundamental physics: the Euler--Heisenberg Lagrangian describes the quantum vacuum response in the low-energy, slowly varying regime of quantum electrodynamics~\cite{Heisenberg:1936nmg}. When coupled to general relativity, such theories provide a framework for determining which properties of charged black holes depend on electromagnetic linearity. Their self-gravitating solutions are particularly useful for distinguishing effects already visible in static configurations from those revealed by rotation, external fields, or nontrivial spacetime geometry.

In four dimensions, the source-free Maxwell equations possess two distinguished symmetries: conformal invariance and continuous electric--magnetic duality. The former preserves the equations under local metric rescalings, whereas the latter rotates the electromagnetic field and its Hodge dual, mixing the equations of motion with the Bianchi identities. Retaining these properties in an interacting theory imposes strong restrictions. Born--Infeld electrodynamics preserves duality, but its dimensionful field scale breaks conformal invariance. Within derivative-free theories of a single Abelian gauge field, imposing both symmetries instead selects a unique nonlinear deformation continuously connected to Maxwell theory. This is ModMax electrodynamics, introduced by Bandos, Lechner, Sorokin, and Townsend~\cite{Bandos:2020jsw}. Its deformation parameter $\gamma$ is dimensionless, and Maxwell theory is recovered at $\gamma=0$. ModMax therefore allows electromagnetic nonlinearities to be studied while retaining both symmetries of the Maxwell equations.

Subsequent developments have explored the structure and applications of this theory. These include supersymmetric extensions~\cite{Bandos:2021rqy}, a six-dimensional conformal chiral two-form counterpart~\cite{Bandos:2020hgy}, and relations to stress-tensor deformation flows~\cite{Ferko:2022iru,Ferko:2022cix}. Exact electromagnetic configurations and their physical consequences have been investigated through knotted fields~\cite{Dassy:2021ulu}, couplings to electric and magnetic charges~\cite{Lechner:2022qhb}, and studies of wave propagation and birefringence~\cite{Neves:2022jqq}. Further work has addressed higher-derivative deformations preserving duality and Weyl invariance~\cite{Kuzenko:2024zra}, Casimir effects~\cite{Sorge:2024gag}, and heat-kernel methods for quantum corrections in nonlinear electrodynamics~\cite{Buchbinder:2026nop}. Holographic applications have also identified coupling-dependent modifications of electric and thermoelectric transport~\cite{Barrientos:2025rde}. {The axial anomaly of charged Dirac fermions has also been studied on Eguchi--Hanson and Taub--NUT instanton backgrounds carrying ModMax fields~\cite{ColipiMarchant:2023AxialAnomaly}.}

Coupling ModMax to Einstein gravity has produced a substantial class of exact four-dimensional solutions. The static dyonic black hole retains the Reissner--Nordstr\"om form, with the nonlinear coupling screening the charge contribution to the metric~\cite{Flores-Alfonso:2020euz}. Dyonic Taub--NUT solutions introduce a nontrivial radial electromagnetic profile~\cite{BallonBordo:2020jtw,Flores-Alfonso:2020nnd}, while the charged C-metric incorporates acceleration, including in the presence of a cosmological constant~\cite{Barrientos:2022bzm}. External fields have been accommodated through ModMax versions of the Melvin--Bonnor universe, its Schwarzschild and C-metric black hole embeddings, and a stationary Melvin--Bonnor--swirling background~\cite{Barrientos:2024umq}. Exact multiblack hole configurations and their dyonic extensions have also been constructed; their extremal geometries retain the Majumdar--Papapetrou form with modified charge-to-mass ratios~\cite{Bokulic:2025usc}. Generalised Harrison transformations in the purely electric and purely magnetic sectors further yield black diholes balanced by external electromagnetic fields~\cite{Bokulic:2025ucy}. More recently, Ernst-type formulations and solution-generating symmetries have been investigated in scalar-coupled models and restricted constitutive sectors~\cite{Bixano:2026ouq,Bixano:2026duf}.

Kerr-like rotation poses a distinct problem. Neither NUT-induced stationarity nor a swirling background provides the angular momentum deformation of an isolated, spherically symmetric charged black hole. More generally, exact rotating solutions in nonlinear electrodynamics are known only in special constructions.\footnote{Garc\'ia D\'iaz obtained exact rotating dyonic solutions supported by a particular nonlinear electrodynamics~\cite{DiazGarcia:2022jpc}. Ay\'on-Beato subsequently determined the supporting theory through explicit structural functions of mixed electromagnetic eigenvalues~\cite{Ayon-Beato:2022dwg}. Although its Lagrangian is not expressible in elementary functions of the usual electromagnetic invariants, the constitutive theory is explicitly defined.} Analytical progress includes slowly rotating solutions in square-root and regularised Maxwell electrodynamics~\cite{Kubiznak:2022vft,Hale:2023dpf}. The latter introduces a dimensionful scale and preserves neither conformal invariance nor continuous electromagnetic duality. Exact near-horizon geometries of extremal rotating Einstein--Born--Infeld black holes provide complementary information, but do not determine the complete rotating exterior~\cite{Hale:2025urg}. These results motivate a perturbative treatment of rotation in ModMax, where both electromagnetic symmetries remain intact.

The difficulty is more specific than the generic complexity of nonlinear field equations. Reference~\cite{Kubiznak:2022vft} analysed slow rotation for parity-even theories $\mathcal L(\mathcal S,\mathcal P^2)$, with electromagnetic invariants defined in Eq.~\eqref{eq:SP}. To illustrate the issue in our notation, the electrically charged ModMax sector at first order in rotation is described by
\begin{align}
 ds^2={}&-f(r)\,dt^2+\frac{dr^2}{f(r)}+r^2d\Omega_2^2-2ar^2\omega(r)\sin^2\theta\,dt\,d\varphi
+\mathcal O(a^2),
 \label{eq:intro_metric}\\
 A={}&-\frac{q}{r}\,dt+aqV(r)\sin^2\theta\,d\varphi+\mathcal O(a^2),
 \label{eq:intro_potential}
\end{align}
with $q$ and $f$ given by Eq.~\eqref{eq:defs} at $Q_m=0$. The functions $\omega$ and $V$ describe frame dragging and the rotation-induced magnetic field, respectively. At $\gamma=0$, the Kerr--Newman solution obeys
\begin{equation}
 \omega(r)=\frac{1-f(r)}{r^2},\qquad V(r)=\frac{1}{r}.
 \label{eq:intro_KN}
\end{equation}
Within the restricted class $\mathcal L(\mathcal S)$, Ref.~\cite{Kubiznak:2022vft} shows that imposing the metric relation in Eq.~\eqref{eq:intro_KN} selects Maxwell electrodynamics. In the broader two-invariant class, the same restriction yields a compatibility condition that ModMax fails for nonzero $\gamma$. This obstructs retaining the Kerr--Newman relation, rather than providing a no-go theorem for rotating ModMax black holes.

The electromagnetic origin of this obstruction is especially transparent for an electrically charged seed. Although the static solution resembles its Maxwell counterpart after charge rescaling, rotation induces a magnetic field and generically produces $\mathcal P=\mathcal O(a)$. Consequently, $\partial\mathcal L/\partial\mathcal P$ contributes to the constitutive equations already at first order in rotation, even though $\mathcal P^2=\mathcal O(a^2)$. The nonlinear response therefore survives when the magnetic monopole charge vanishes. A charge-rescaled Kerr--Newman potential cannot capture this response: its magnetic profile must be determined together with the frame-dragging function. We accordingly retain $\omega(r)$ and $V(r)$ as independent functions fixed by the coupled field equations.

{This departure from Kerr--Newman has a simple physical origin, already visible in flat spacetime. Around a nonzero electric Coulomb background, the linearised magnetic constitutive response of ModMax distinguishes radial and tangential perturbations, with coefficients that remain different even as the background electric field becomes arbitrarily weak. As a consequence, the magnetic dipole exhibits an anomalous fractional-power decay, $V(r)\sim r^{-\delta}$, where $\delta=(\sqrt{1+8e^{-2\gamma}}-1)/2$, instead of the Maxwell behaviour $V(r)\sim 1/r$. This exponent belongs to the anomalous magnetic multipole spectrum previously identified for perturbations around electric ModMax backgrounds~\cite{BeltranJimenezBettoniBrax:2023}. Its origin lies in the nonlinear electromagnetic constitutive law rather than in gravity, providing an asymptotic explanation for the incompatibility of Kerr--Newman-type magnetic profiles with the rotating Einstein--ModMax equations.}

Our aforementioned choice provides a controlled route to rotational deformations of the static Einstein--ModMax black hole. At first order in $a$, the angular dependence separates, and the problem reduces to coupled radial equations, while finite rotation requires a more general treatment of the radial and angular dependence. We construct new analytic solutions on the branch connected to Kerr--Newman as $\gamma\to0$, using joint expansions in $a$ and $\gamma$ to resolve their coupled gravitational and electromagnetic corrections. The perturbative coefficients satisfy the field equations exactly at the retained orders; the resulting expressions approximate the finite-rotation geometry. We also retain arbitrary $\gamma$ at first order in $a$ and develop a convergent series representation. Together, these constructions identify the departures from the Kerr--Newman profiles required by the nonlinear constitutive law and provide a basis for investigating the fully rotating problem. {In particular, the arbitrary-coupling solution exhibits the anomalous magnetic falloff dictated by the flat-space constitutive response, while horizon regularity determines its amplitude and gravitational backreaction.}

The paper is organised as follows. \autoref{sec2} presents Einstein--ModMax theory, its field equations, and the metric and electromagnetic ans\"atze. \autoref{sec:solutions} constructs the solutions through the mixed orders $\mathcal O(a\gamma)$, $\mathcal O(a\gamma^2)$, and $\mathcal O(a^2\gamma)$, retaining the lower-order terms in each expansion, and develops the series solution at first order in $a$ with arbitrary $\gamma$. \autoref{sec:final_remarks} concludes with final remarks and directions for further investigation.

\section{Preliminaries}\label{sec2}

\subsection{Action, field equations and charge conventions}
\label{sec:action}

We work in four spacetime dimensions, adopt the metric signature $(-,+,+,+)$, and use units in which $c=G=1$. With vanishing cosmological constant, Einstein--ModMax theory is governed by the action
\begin{equation}
 I[g,A]=\frac{1}{16\pi}\int d^4x\,\sqrt{-g}\,\bigl[R+4\mathcal L(\mathcal S,\mathcal P)\bigr],
 \label{eq:action}
\end{equation}
where $R$ is the Ricci scalar and
\begin{equation}
 \mathcal L(\mathcal S,\mathcal P)=-\mathcal S\cosh\gamma+\sqrt{\mathcal S^2+\mathcal P^2}\sinh\gamma.
 \label{eq:L}
\end{equation}
The dimensionless parameter $\gamma$ controls the nonlinear deformation, and we restrict attention to $\gamma\geq0$. Maxwell electrodynamics is recovered at $\gamma=0$, for which $\mathcal L=-\mathcal S$. The electromagnetic invariants are defined by
\begin{equation}
 F=dA,\qquad
 \mathcal S=\frac14 F_{\mu\nu}F^{\mu\nu},\qquad \mathcal P=\frac14 F_{\mu\nu}({}^{\star}F)^{\mu\nu}.
 \label{eq:SP}
\end{equation}
Our Hodge-dual convention is
\begin{equation}
 ({}^{\star}F)_{\mu\nu}
=\frac12\epsilon_{\mu\nu\rho\sigma}F^{\rho\sigma},\qquad\epsilon_{tr\theta\varphi}=+\sqrt{-g},
\end{equation}
so that ${}^{\star}{}^{\star}F=-F$.

The nonlinear constitutive relation is conveniently expressed in terms of the excitation two-form
\begin{equation}
 H_{\mu\nu}=-\mathcal L_{\mathcal S}F_{\mu\nu}-\mathcal L_{\mathcal P}({}^{\star}F)_{\mu\nu},
\end{equation}
where subscripts on $\mathcal L$ denote partial derivatives. In the non-null sector $\mathcal S^2+\mathcal P^2\neq0$, this becomes
\begin{equation}
 \begin{aligned}
 H_{\mu\nu}
 =\mathcal C F_{\mu\nu}
 -\frac{\mathcal P\sinh\gamma}
 {\sqrt{\mathcal S^2+\mathcal P^2}}
 ({}^{\star}F)_{\mu\nu},
 \end{aligned}
 \label{eq:CH}
\end{equation}
where we have defined
\begin{equation}
    \mathcal C=\cosh\gamma-\frac{\mathcal S\sinh\gamma}{\sqrt{\mathcal S^2+\mathcal P^2}}.\quad
\end{equation}
{For strictly electric configurations,
$\mathcal S<0$ and $\mathcal P=0$, the constitutive
relation reduces to $H_{\mu\nu}=e^\gamma F_{\mu\nu}$,
whereas for strictly magnetic configurations,
$\mathcal S>0$ and $\mathcal P=0$, it becomes
$H_{\mu\nu}=e^{-\gamma}F_{\mu\nu}$. Thus, each
sector separately behaves as Maxwell electrodynamics
with a rescaled coupling. This simplification does
not extend to perturbations that mix electric and
magnetic fields: around an electric background,
the term proportional to $\mathcal P$ in
Eq.~\eqref{eq:CH} contributes already
at first order in a magnetic perturbation.}

Variation of the action with respect to the metric and gauge potential, together with the Bianchi identity, yields
\begin{equation}
 \begin{aligned}
 G_{\mu\nu}
=2\mathcal C
\bigl(F_{\mu\rho}F_\nu{}^\rho-\mathcal Sg_{\mu\nu}\bigr),\quad\nabla_\mu H^{\mu\nu}=0,\qquad\nabla_\mu({}^{\star}F)^{\mu\nu}=0.
 \end{aligned}
 \label{eq:EinsteinModMAx}
\end{equation}
The compact form of the Einstein equation follows from the homogeneity identity
$\mathcal L=\mathcal S\mathcal L_{\mathcal S}+\mathcal P\mathcal L_{\mathcal P}$.
In the Maxwell limit, $\mathcal C=1$ and $H=F$, recovering the standard Einstein--Maxwell equations.

Finally, we define the conserved electric and magnetic charges by
\begin{equation}
 Q_e=\frac{1}{4\pi}\int_{S^2_\infty}{}^{\star}H, \qquad Q_m=\frac{1}{4\pi}\int_{S^2_\infty}F.
 \label{eq:charges}
\end{equation}
The flux of the nonlinear excitation therefore determines the electric charge, while the magnetic charge is the flux of the field strength.

\subsection{Metric and electromagnetic ans\"atze and perturbative organisation}
\label{sec:ansatz}

We employ complementary expansions in the rotation parameter $a$ and the ModMax coupling $\gamma$, holding the physical parameters $M$, $Q_e$, and $Q_m$ fixed. Throughout, an expansion through $\mathcal O(a^m\gamma^n)$ retains all terms $a^i\gamma^j$ with $0\leq i\leq m$ and $0\leq j\leq n$. The three finite truncations considered below correspond to $(m,n)=(1,1)$, $(1,2)$, and $(2,1)$. We also consider the sector linear in $a$ without expanding in $\gamma$. For compactness, the exactly known static dependence on $\gamma$ may be retained unexpanded; this does not extend the accuracy claimed for the rotational corrections.

\subsubsection{Linear rotation at arbitrary coupling}

As discussed in the introduction, the Kerr--Newman restriction $\omega=(1-f)/r^2$ is incompatible with the charged ModMax equations for $\gamma\neq0$~\cite{Kubiznak:2022vft}. We therefore retain independent frame-dragging and magnetic profiles, $\omega(r)$ and $V(r)$.

Define
\begin{equation}
 \begin{aligned}
 k&=Q_e^2+Q_m^2,& K&=e^{-\gamma}k,\\
 q&=e^{-\gamma}Q_e,&
 f(r)&=1-\frac{2M}{r}+\frac{K}{r^2}.
 \end{aligned}
 \label{eq:defs}
\end{equation}
Here $q$ is the electrostatic amplitude, whereas $Q_e$ is the physical excitation charge defined in Eq.~\eqref{eq:charges}. We consider charged, nonextremal seeds with $M>0$ and $0<K<M^2$, and normalise $a=J/M$. We also fix the static horizon notation used throughout:
\begin{equation}
 d=\sqrt{M^2-K},\qquad r_\pm=M\pm d,\qquad
 \Delta(r)=r^2f(r)=(r-r_+)(r-r_-).
 \label{eq:static_horizons}
\end{equation}
The metric is
\begin{equation}
 ds^2=-f\,dt^2+\frac{dr^2}{f}+r^2\bigl(d\theta^2+\sin^2\theta\,d\varphi^2\bigr)-2ar^2\omega\sin^2\theta\,dt\,d\varphi+\mathcal O(a^2).
 \label{eq:metric}
\end{equation}
Only the off-diagonal metric component changes at this order. The corresponding electromagnetic ansatz is
\begin{equation}
 \begin{aligned}
 A={}&\left[-\frac{q}{r}+aQ_m\bigl(\omega-fV'\bigr)\cos\theta\right]dt+\left[-Q_m\cos\theta+aqV\sin^2\theta\right]d\varphi+\mathcal O(a^2),
 \end{aligned}
 \label{eq:Adyonexact}
\end{equation}
where primes denote radial derivatives. This form reflects electromagnetic duality: the electric dipole associated with $Q_m$ is determined by the same profiles that govern frame dragging and the magnetic dipole. Consequently, this family requires no additional radial function for the dyonic extension. The monopole potential is understood in two patches, $A_N=A+Q_m\,d\varphi$ and $A_S=A-Q_m\,d\varphi$.

At $\gamma=0$, the choices $V=1/r$ and $\omega=(1-f)/r^2$ recover the dyonic Kerr--Newman solution through first order in $a$. At finite $\gamma$, both profiles are determined by the coupled equations. Equations~\eqref{eq:metric} and~\eqref{eq:Adyonexact} also provide the starting point for the exterior series construction at arbitrary coupling.

{Importantly, the ansatz does not
impose a Maxwell-like $1/r$ decay on the rotational
magnetic profile $V(r)$. Instead, its asymptotic
behaviour is determined by the field equations,
with the growing electromagnetic mode excluded
at infinity. As shown below, the decaying mode
generically exhibits an anomalous fractional-power
falloff for nonzero $\gamma$.}

In the coupling expansions below, retaining $K$ unexpanded preserves the exact static geometry and its horizon position while keeping the rotational corrections compact.
This amounts to a partial resummation of the coupling expansion. Since the radial equations are satisfied to the required order for arbitrary admissible $K$, restoring its coupling dependence preserves that accuracy. Expanding all implicit coupling dependence recovers the strict Taylor expansion at fixed physical charges. No accuracy beyond the stated rotational orders is implied. A regular expansion about $\gamma=0$ assumes $0<k<M^2$; the extremal limit requires a separate treatment.

\subsubsection{Linear rotation through first and second order in the coupling}

The sectors $\mathcal O(a\gamma)$ and $\mathcal O(a\gamma^2)$ require no new angular structures. They follow from the preceding ansatz by expanding its two radial profiles. Introduce the reference function
\begin{equation}
 \omega_{\mathrm s}(r;K)=\frac{1-f(r)}{r^2}=\frac{2M}{r^3}-\frac{K}{r^4},
\label{eq:reference_dragging}
\end{equation}
and write, for $N=1,2$,
\begin{equation}
 \begin{aligned}
 V_{[N]}(r)&=\frac1r+\sum_{n=1}^{N}\gamma^n v_n(r;K),\\
 \omega_{[N]}(r)&=\omega_{\mathrm s}(r;K)+\sum_{n=1}^{N}\gamma^n w_n(r;K).
 \end{aligned}
\label{eq:profiles_gamma_expansion}
\end{equation}
Thus $N=1$ gives the $a\gamma$ construction, while $N=2$ adds the $a\gamma^2$ corrections. Substituting these expressions into Eqs.~\eqref{eq:metric} and~\eqref{eq:Adyonexact} specifies both configurations without repeating their tensorial form. In particular, the electric dipole profile becomes
\begin{equation}
 \omega_{[N]}-fV_{[N]}'=\frac1{r^2}+\sum_{n=1}^{N}\gamma^n\bigl(w_n-fv_n'\bigr).
\label{eq:dipole_gamma_expansion}
\end{equation}
The rotational remainders are $\mathcal O(a^2)$ and $\mathcal O(a\gamma^{N+1})$, while the static sector remains exact.

\subsubsection{Quadratic rotation at first order in the coupling}

For the $\mathcal O(a^2\gamma)$ construction, we restrict Eq.~\eqref{eq:defs} to the electric sector, $Q_m=0$. The functions $v_1$ and $w_1$ are precisely those introduced in Eq.~\eqref{eq:profiles_gamma_expansion}; they are not new unknown profiles. At second order in rotation, the diagonal metric and electrostatic potential acquire additional angular dependence. The quadratic sources contain monopole and quadrupole harmonics, whereas the frame-dragging and azimuthal-potential corrections remain odd in $a$.

To incorporate the known Maxwell contributions without duplication, introduce the Kerr--Newman algebraic expressions with screened charge parameter $K$,
\begin{equation}
 \Sigma=r^2+a^2\mu^2,\qquad \Delta_a=\Delta+a^2,\qquad \mu=\cos\theta,\quad s^2=1-\mu^2,
\label{eq:screened_KN_parameters}
\end{equation}
\begin{equation}
 \begin{aligned}
 ds^2_{\mathrm{sKN}}={}&-\frac{\Delta_a}{\Sigma}(dt-as^2d\varphi)^2+\frac{\Sigma}{\Delta_a}\,dr^2+\Sigma\,d\theta^2+\frac{s^2}{\Sigma}\bigl[(r^2+a^2)d\varphi-a\,dt\bigr]^2,\\
 A_{\mathrm{sKN}}={}&-\frac{qr}{\Sigma}(dt-as^2d\varphi).
 \end{aligned}
 \label{eq:screened_KN_seed}
\end{equation}
These expressions serve only as a reference configuration; they do not solve the rotating ModMax equations when $\gamma\neq0$. In the formulas below, $[\cdots]_{\leq a^2}$ denotes retaining terms through second order in $a$.

The additional even-parity fields are parametrised by
\begin{equation}
 \begin{aligned}
U(r,\theta)&=u_0(r)+u_2(r)P_2(\mu),\\
B(r,\theta)&=b_0(r)+b_2(r)P_2(\mu),\\
\nu(r,\theta)&=\frac{u_0(r)+u_2(r)}2-\frac{\mathcal Z(r)}2s^2,\qquad P_2(\mu)=\frac12(3\mu^2-1).
 \end{aligned}
\label{eq:second_order_harmonics}
\end{equation}
The parametrisation of $\nu$ implements elementary flatness on the axis, where $2\nu=U$. In a circular coordinate gauge adapted to the Killing and orbit-space blocks, the metric and potential take the form
 \begin{align}
 ds^2={}&[ds^2_{\mathrm{sKN}}]_{\leq a^2}-2\gamma a r^2w_1s^2\,dt\,d\varphi+\gamma a^2\bigg[\left(fU+2r^2\omega_{\mathrm s}w_1s^2-\frac{\tau}{Mr}\right)dt^2\nonumber\\
 &+\left(\frac{2\nu}{f}-\frac{\tau}{Mrf^2}\right)dr^2+2r^2\nu\,d\theta^2+r^2s^2U\,d\varphi^2\bigg]+\mathcal O(a^3,a\gamma^2),\label{eq:metric_a2gamma}\\
 A={}&[A_{\mathrm{sKN}}]_{\leq a^2}+q\gamma a v_1s^2\,d\varphi+q\gamma a^2B\,dt+\mathcal O(a^3,a\gamma^2).
\label{eq:potential_a2gamma}
 \end{align}
The term proportional to $\omega_{\mathrm s}w_1$ captures the first-order frame-dragging contribution to the second-order diagonal geometry. The electric monopole and quadrupole profiles $b_0,b_2$ are additional radial unknowns, determined by second-order equations whose sources are fixed by $v_1$ and $w_1$.

The terms proportional to $\tau$ implement a homogeneous mass adjustment, fixed so that $M$ remains the ADM mass. Before this adjustment, the parametrisation obeys
\begin{equation}
 \det(G_{AB})=-\Delta_a\sin^2\theta,\qquad
 G_{AB}=g_{AB},\qquad A,B\in\{t,\varphi\},
\label{eq:intermediate_determinant}
\end{equation}
through the retained orders. This condition fixes the determinant of the Killing block in the intermediate gauge; the final mass-normalised metric need not preserve it. Asymptotic normalisation fixes the constant $\tau$, which introduces no additional physical parameter. Likewise, the remaining Einstein constraints determine $\mathcal Z$ once the coupled profiles $u_0,u_2,b_0,b_2$ have been obtained.

Equations~\eqref{eq:metric_a2gamma} and~\eqref{eq:potential_a2gamma} include the lower-order electric solution and the Maxwell terms through $a^2$. Their new content is the even-parity correction at $a^2\gamma$. The complete $a\gamma^2$ corrections determined in the preceding sector are not retained in this truncation. Thus, the three perturbative sectors share a common pair of rotational profiles and, at quadratic order in $a$ only, the additional monopole and quadrupole fields.

\section{Analytic rotating solutions}
\label{sec:solutions}

\subsection{Linear rotation through first order in the coupling}
\label{sec:agamma}

We begin with the $\mathcal O(a\gamma)$ sector, corresponding to $N=1$ in Eq.~\eqref{eq:profiles_gamma_expansion}. We first construct the electrically charged solution and then obtain its dyonic extension by electromagnetic duality. Throughout this subsection, we retain the static functions and their dependence on $K$ as specified in \autoref{sec:ansatz}; the omitted rotational terms are $\mathcal O(a^2)$ and $\mathcal O(a\gamma^2)$.

\paragraph{Electric solution.}

We set $Q_m=0$ in Eq.~\eqref{eq:defs}. Substituting the first-order rotational ansatz into the field equations, before expanding in $\gamma$, gives
\begin{equation}
 (fV')'-\frac{2e^{-2\gamma}}{r^2}V=\omega', \qquad (r^4\omega')'=4KV'.
 \label{eq:agamma_coupled}
\end{equation}
The integration constant in the Einstein equation is fixed by $J=aM$. In a nonrotating frame at infinity,
\begin{equation}
 r^4\omega'=4KV-6M,
 \qquad \omega(\infty)=0.
\label{eq:agamma_spin_normalisation}
\end{equation}
Writing
\begin{equation}
 V=\frac1r+\gamma v_1(r;K), \qquad \omega=\omega_{\mathrm s}+\gamma w_1(r;K),
 \label{eq:agamma_profiles}
\end{equation}
with $\omega_{\mathrm s}$ defined in Eq.~\eqref{eq:reference_dragging}, and retaining first order in the explicit coupling gives
\begin{equation}
 \mathcal L_K v_1\equiv(fv_1')' -\left(\frac2{r^2}+\frac{4K}{r^4}\right)v_1=-\frac4{r^3},
 \qquad
 w_1'=\frac{4K}{r^4}v_1.
 \label{eq:agamma_reduced}
\end{equation}
In particular, $w_1$ contains only the additional rotational correction: the static charge screening is already included in the reference term $-K/r^4$.

{The source term $-4/r^3$ in the equation for $v_1$ originates from the coupling-dependent coefficient $e^{-2\gamma}$ in the magnetic field equation. It therefore reflects the nontrivial constitutive response to a magnetic perturbation around the electric background, rather than the screening of the charge contribution to the static metric.}

Using the static horizon quantities in Eq.~\eqref{eq:static_horizons}, the integration of Eq.~\eqref{eq:agamma_reduced} simplifies because
\begin{equation}
 \Pi(r)=r^3-3Kr+\frac{2K^2}{M},
 \qquad y(r)=\frac{\Pi(r)}r,
 \qquad \mathcal L_K y=0.
\label{eq:agamma_homogeneous}
\end{equation}
The associated Lagrange identity can therefore be integrated once as
\begin{align}
 f\bigl(yv_1'-y'v_1\bigr)&=N(r),
\label{eq:agamma_first_integral}\\
 N(r)&=-4\log\frac r{r_+}
 -6K\left(\frac1{r^2}-\frac1{r_+^2}\right)
 +\frac{8K^2}{3M}\left(\frac1{r^3}-\frac1{r_+^3}\right).
 \label{eq:agamma_N}
\end{align}
Here the integration constant has been chosen so that $N(r_+)=0$, as required by horizon regularity. The remaining homogeneous contribution is fixed by excluding the growing electromagnetic mode at infinity.

To give the resulting solution without unevaluated integrals, introduce
\begin{align}
 \mathcal D(r)={}&
\operatorname{Li}_2\!\left(\frac{r_+}{r}\right) -\operatorname{Li}_2\!\left(\frac{r_-}{r}\right) -\log\frac r{r_+}\log\frac{r-r_+}{r-r_-},
 \label{eq:agamma_D}\\
 \mathcal A(r)={}&\frac{3M^2r^2+3M^3r-2K(M^2+2K)}{12},
 \label{eq:agamma_Aaux}\\
 \mathcal B(r)={}&-\frac{5M^2}{6r}-\frac{M(3M^2+2K)}{6r^2}+\frac{2K(M^2+2K)}{9r^3}+\frac{M^2}{2d}\mathcal D(r)+\frac{M(M^2+2K)}{3K}\log\left(1-\frac{r_-}{r}\right),
 \label{eq:agamma_Baux}
\end{align}
where $\operatorname{Li}_2$ denotes the dilogarithm. All functions are taken on their real branches in the exterior region $r>r_+$. The two radial corrections are
\begin{align}
 v_1(r;K)&=
 \frac{\mathcal A(r)N(r)+\Pi(r)\mathcal B(r)}{r d^4},
 \label{eq:agamma_v_closed}\\
 w_1(r;K)&=
 \frac{2K}{3M\Pi(r)}
 \left[\frac{3f(r)(r^2-K)}r\,v_1(r;K)+N(r)\right]
 -\frac{8K}{9Mr^3}.
 \label{eq:agamma_w_closed}
\end{align}
Thus both profiles are finite expressions in rational functions, logarithms, and dilogarithms, with no undetermined integration constants. The identities
\begin{equation}
 \mathcal D'=-\log\frac r{r_+}
 \left(\frac1{r-r_+}-\frac1{r-r_-}\right),
 \qquad N'=-\frac{4\Pi}{r^4},
\label{eq:agamma_derivative_identities}
\end{equation}
allow direct verification of Eq.~\eqref{eq:agamma_first_integral} and of the reduced field equations~\eqref{eq:agamma_reduced}.

The electric solution is consequently
\begin{align}
 ds^2={}&-f\,dt^2+\frac{dr^2}{f}+r^2\bigl(d\theta^2+\sin^2\theta\,d\varphi^2\bigr)-2ar^2\left(\omega_{\mathrm s}+\gamma w_1\right)\sin^2\theta\,dt\,d\varphi+\mathcal O(a^2,a\gamma^2),
 \label{eq:agamma_metric}\\
 A={}&-\frac q r\,dt+aq\left(\frac1r+\gamma v_1\right)\sin^2\theta\,d\varphi+\mathcal O(a^2,a\gamma^2).
\label{eq:agamma_electric_potential}
\end{align}
The correction $v_1$ resolves the failure of the screened Kerr--Newman magnetic profile, while $w_1$ accounts for its gravitational backreaction.

\paragraph{Dyonic extension.}

Electromagnetic duality generates the solution with both charges from an electric representative of charge $\sqrt{Q_e^2+Q_m^2}$. The stress tensor is invariant under this transformation, so the line element remains Eq.~\eqref{eq:agamma_metric}, with the general dyonic parameters of Eq.~\eqref{eq:defs}. Every auxiliary function and radial correction above is evaluated with the corresponding $K$. No additional radial equation is required.

For an explicit expression for the induced electric dipole, define
\begin{align}
 \Xi_1(r;K)&=w_1-fv_1'
 \nonumber\\
 &=\frac{2rf(r)(K/M-r)v_1(r;K)
 +(2K/(3M)-r)N(r)}{\Pi(r)}
 -\frac{8K}{9Mr^3}.
 \label{eq:agamma_Xi_closed}
\end{align}
The dyonic potential then takes the closed form
\begin{equation}
 \begin{aligned}
 A={}&\left[-\frac q r+aQ_m\left(\frac1{r^2}+\gamma\Xi_1\right)\cos\theta\right]dt+\left[-Q_m\cos\theta+aq\left(\frac1r+\gamma v_1\right)\sin^2\theta\right]d\varphi+\mathcal O(a^2,a\gamma^2).
 \end{aligned}
\label{eq:agamma_dyonic_potential}
\end{equation}
This is the $N=1$ realisation of Eq.~\eqref{eq:Adyonexact}, since $\omega-fV'=r^{-2}+\gamma\Xi_1$. The monopole term is understood in the two regular gauge patches specified in \autoref{sec:ansatz}. At $\gamma=0$, the configuration reduces to the dyonic Kerr--Newman solution through first order in $a$.

\paragraph{Regularity and asymptotic behaviour.}

The apparently singular logarithmic product in Eq.~\eqref{eq:agamma_D} has a regular outer-horizon limit,
\begin{equation}
 \mathcal D(r_+)=\frac{\pi^2}{6}
 -\operatorname{Li}_2\!\left(\frac{r_-}{r_+}\right).
 \label{eq:agamma_D_horizon}
\end{equation}
Moreover, $\Pi(r_+)=2r_+^2d^2/M>0$ and $\Pi'=3(r^2-K)>0$ throughout the exterior. The displayed denominators therefore introduce no exterior poles, and $v_1$, $w_1$, and $\Xi_1$ are regular at $r_+$. With $\Omega_H=a\omega(r_+)$ at the retained order, the angular dependence of $A_t+\Omega_H A_\varphi$ cancels on the horizon. Its value is $-q/r_+$ in the displayed monopole gauge; the regular patches add only angle-independent constants.

For completeness, introduce
\begin{equation}
 \ell_r=\frac43\log\frac r{r_+}+c_\infty,
 \qquad
 c_\infty=\frac49-\frac{2K}{r_+^2}
 +\frac{8K^2}{9Mr_+^3}.
\label{eq:agamma_asymptotic_coefficient}
\end{equation}
The radial corrections behave as
\begin{align}
 v_1&=\frac{\ell_r}{r}+\mathcal O\!\left(\frac{\log(r/r_+)}{r^2}\right),
 \nonumber\\
 w_1&=-\frac{K}{r^4}\left(\ell_r+\frac13\right)+\mathcal O\!\left(\frac{\log(r/r_+)}{r^5}\right),\nonumber\\
 \Xi_1&=\frac{\ell_r-4/3}{r^2}+\mathcal O\!\left(\frac{\log(r/r_+)}{r^3}\right).
 \label{eq:agamma_asymptotics}
\end{align}

{The logarithmic term in the magnetic profile has a direct physical interpretation. As discussed in the Introduction and further detailed in the Appendix, the magnetic dipole generated around an electric Coulomb background in ModMax electrodynamics exhibits an anomalous radial falloff, $V\sim r^{-\delta}$, with $\delta=1-\frac{4}{3}\gamma+\mathcal O(\gamma^2)$. The logarithm in $v_1$ is precisely the first-order expansion of this fractional power. In particular, the correction cannot be interpreted as a finite shift of the conventional Maxwell magnetic dipole moment. The asymptotic magnetic response is instead characterised by a modified radial scaling, whose complete finite-coupling form we establish below.}

The rotational electromagnetic corrections therefore decay, despite their logarithmic factors. In the dyonic case, the surviving $-Q_m\cos\theta\,d\varphi$ term is the usual monopole connection, whose physical magnetic field falls off as $r^{-2}$. The leading metric asymptotics remain $g_{tt}=-1+2M/r+\mathcal O(r^{-2})$ and $g_{t\varphi}=-2J\sin^2\theta/r+o(r^{-1})$, establishing asymptotic flatness with the prescribed mass and angular momentum. These expressions solve the coupled equations through the retained order.

{Although the electromagnetic corrections decay at each retained order, the small-coupling expansion of the magnetic profile is not uniform at spatial infinity. Its relative validity requires $|\gamma\log(r/r_+)|\ll1$. Consequently, the asymptotic magnetic moment cannot be extracted by taking $r\to\infty$ at fixed order in $\gamma$. The finite-coupling construction below resums the logarithmic behaviour into the anomalous fractional power.} The extremal limit requires a separate analysis.

\subsection{Linear rotation through second order in the coupling}
\label{sec:agamma2}

We now extend the electric solution to $\mathcal O(a\gamma^2)$, using Eq.~\eqref{eq:defs} with $Q_m=0$. The $N=2$ specialisation of Eq.~\eqref{eq:profiles_gamma_expansion} is
\begin{equation}
 \begin{aligned}
 V&=\frac1r+\gamma v_1(r;K)+\gamma^2v_2(r;K)+\mathcal O(\gamma^3),\\
 \omega&=\omega_{\mathrm s}+\gamma w_1(r;K)+\gamma^2w_2(r;K)+\mathcal O(\gamma^3).
 \end{aligned}
 \label{eq:agamma2_profiles}
\end{equation}
The reference profile is given in Eq.~\eqref{eq:reference_dragging}, and $v_1,w_1$ are those constructed in \autoref{sec:agamma}; no new angular structures are required. The screened parameter $K$ remains unexpanded.

Expanding the radial equations~\eqref{eq:agamma_coupled} through second order in the explicit coupling gives
\begin{equation}
 \mathcal L_Kv_2=\frac4{r^3}-\frac{4v_1}{r^2},
 \qquad
 w_2'=\frac{4K}{r^4}v_2,
\label{eq:agamma2_radial_equations}
\end{equation}
where $\mathcal L_K$ is defined in Eq.~\eqref{eq:agamma_reduced}. Thus the second-order magnetic correction is sourced entirely by the known first-order profile. As before, horizon regularity and the exclusion of a growing electromagnetic mode fix $v_2$, while the normalisations $J=aM$ and $\omega(\infty)=0$ fix $w_2$.

{The two contributions to the source of $v_2$ arise from the expansion of the coupling-dependent constitutive coefficient $e^{-2\gamma}$: the term $4/r^3$ is generated at second order, while $-4v_1/r^2$ accounts for its action on the first-order magnetic correction. The successive coupling corrections thus reconstruct the same modified magnetic response identified in \autoref{sec:agamma}, rather than introducing a new asymptotic multipole structure.}

\paragraph{Dimensionless radial problem.}

Introduce
\begin{equation}
 x=\frac{r_+}{r},\qquad
 \rho=\frac{r_-}{r_+},\qquad
 0<\rho<1,\qquad
 \widehat v_n(x)=r_+v_n(r_+/x;K).
 \label{eq:agamma2_variables}
\end{equation}
Here $r_\pm$ are given in Eq.~\eqref{eq:static_horizons}, so that $K=\rho r_+^2$ and $2M=(1+\rho)r_+$. Infinity corresponds to $x=0$ and the outer horizon to $x=1$. Define the dimensionless operator
\begin{equation}
 \mathcal D_\rho u=\partial_x\!\left[p(x)\partial_xu\right]-(2+4\rho x^2)u,
 \qquad p(x)=x^2(1-x)(1-\rho x).
 \label{eq:agamma2_operator}
\end{equation}
The perturbative equations become
\begin{equation}
 \mathcal D_\rho \widehat v_1=-4x,\qquad
 \mathcal D_\rho \widehat v_2=g_2(x),\qquad
 g_2(x)=4x-4\widehat v_1(x).
 \label{eq:agamma2_sources}
\end{equation}
In particular, $\widehat v_1$ is explicitly known from Eq.~\eqref{eq:agamma_v_closed}.

An elementary homogeneous basis is
\begin{align}
 Y_{\mathrm H}(x)&=\frac1{x^2}-3\rho+\frac{4\rho^2}{1+\rho}x,
 \label{eq:agamma2_YH}\\
 Y_\infty(x)&=\frac{1}{(1-\rho)^4}
 \left[-\frac{(1+\rho)^2}{x}-\frac{(1+\rho)^3}{2}
 +\frac{2\rho(\rho^2+10\rho+1)}{3}x\right]
 +\frac{(1+\rho)^2}{(1-\rho)^5}
 Y_{\mathrm H}(x)\log\frac{1-\rho x}{1-x}.
 \label{eq:agamma2_Yinf}
\end{align}
These functions satisfy
\begin{equation}
 \mathcal D_\rho Y_{\mathrm H}
 =\mathcal D_\rho Y_\infty=0,\qquad
 p\left(Y_{\mathrm H}\partial_xY_\infty
 -Y_\infty\partial_xY_{\mathrm H}\right)=1.
 \label{eq:agamma2_Wronskian}
\end{equation}
The mode $Y_{\mathrm H}$ is regular at the horizon and grows as $x^{-2}$ at infinity, whereas $Y_\infty=x/3+\mathcal O(x^2)$ decays at infinity and has a logarithmic singularity at the horizon.

\paragraph{Finite polylogarithmic representation.}

The inhomogeneous solution can be expressed using Goncharov multiple polylogarithms (GPL) with alphabet
\begin{equation}
 \mathcal A_\rho=\{0,1,\rho^{-1}\}.
 \label{eq:agamma2_alphabet}
\end{equation}
We use the base-point-zero convention
\begin{equation}
 \partial_xG(z,\boldsymbol w;x)
 =\frac{G(\boldsymbol w;x)}{x-z},\qquad
 G(\varnothing;x)=1,\qquad
 G(\underbrace{0,\ldots,0}_{n};x)=\frac{\log^n x}{n!}.
\label{eq:agamma2_GPL_definition}
\end{equation}
For example, $\log(1-zx)=G(z^{-1};x)$ and $\operatorname{Li}_2(zx)=-G(0,z^{-1};x)$ for $z=1,\rho$.

To keep the result compact, let $\mathfrak P[F](x)$ denote the antiderivative of $F$ whose constant finite part at $x=0$ vanishes. The finite part is the coefficient of $x^0\log^0x$ in the local power-log expansion. This prescription is an algebraic operation on GPL expressions: after expanding products by the shuffle identity and taking partial fractions, use
\begin{align}
 \mathfrak P\!\left[\frac{G(\boldsymbol w;x)}{x-z}\right]
 &=G(z,\boldsymbol w;x),
\label{eq:agamma2_primitive_simple}\\
 \mathfrak P\!\left[x^nG(z,\boldsymbol w;x)\right]
 &=\operatorname{Reg}_0\!\left[
 \frac{x^{n+1}G(z,\boldsymbol w;x)}{n+1}
 -\frac1{n+1}
 \mathfrak P\!\left[
 \frac{x^{n+1}}{x-z}G(\boldsymbol w;x)\right]\right],
 \quad n\in\mathbb Z,\quad n\ne-1,
\label{eq:agamma2_primitive_power}
\end{align}
where $\operatorname{Reg}_0F=F-\operatorname{FP}_{x=0}F$. For the empty word, the elementary rules are $\mathfrak P[x^n]=x^{n+1}/(n+1)$ for $n\ne-1$ and $\mathfrak P[x^{-1}]=\log x$. All rational coefficients entering the construction below are Laurent polynomials in $x$; subsequent partial fractions require only Laurent powers and simple poles at the letters of $\mathcal A_\rho$. Each integration-by-parts step shortens the GPL word, so these rules give a finite reduction.

Define
\begin{equation}
 C_2(x)=\mathfrak P[Y_{\mathrm H}g_2](x),
 \label{eq:agamma2_C2}
\end{equation}
and set
\begin{equation}
 \widehat v_2(x)=Y_\infty(x)\bigl[C_2(x)-C_2(1)\bigr]
 -Y_{\mathrm H}(x)\mathfrak P[Y_\infty g_2](x).
 \label{eq:agamma2_u2_closed}
\end{equation}
The required radial corrections are
\begin{equation}
 v_2(r;K)=\frac{\widehat v_2(x)}{r_+},
 \qquad
 w_2(r;K)=-\frac{4\rho}{r_+^2}
 \mathfrak P[x^2\widehat v_2](x).
\label{eq:agamma2_radial_solution}
\end{equation}
With $x$ as in Eq.~\eqref{eq:agamma2_variables}, these equations and the explicit first-order profile provide a finite GPL representation of both corrections. The endpoint value $C_2(1)$ is fixed by the displayed functions and is not an additional integration constant.

The first-order function $\widehat v_1$ contains polylogarithms of weight at most two. The construction produces $\widehat v_2$ of weight at most four. Its weight-four terms carry the rational prefactor $Y_{\mathrm H}$; since $x^2Y_{\mathrm H}$ is polynomial, the primitive defining $w_2$ also has weight at most four. Thus both new profiles remain within the same finite GPL class, without introducing additional letters. The expressions above specify the finite reduction rather than listing every GPL word separately.

\paragraph{Boundary conditions and verification.}

At infinity, $g_2=\mathcal O(x\log x)$, so $\mathfrak P[Y_\infty g_2]=\mathcal O(x^3\log x)$, while $C_2$ contains only logarithmic divergences. Consequently,
\begin{equation}
 \widehat v_2=\mathcal O(x\log^2x),\qquad
 \mathfrak P[x^2\widehat v_2]=\mathcal O(x^4\log^2x),
 \label{eq:agamma2_infinity}
\end{equation}
which excludes the growing electromagnetic mode and preserves the mass and angular momentum normalisations.

{
The leading logarithmic terms can be determined directly from the radial equation, independently of the full GPL representation. Using the asymptotic behaviour
$\widehat v_1(x)=x[c_\infty-\frac43\log x]+\mathcal O(x^2\log x)$, with $c_\infty$ defined in Eq.~\eqref{eq:agamma_asymptotic_coefficient}, one obtains
\begin{equation}
 \widehat v_2(x)
 =x\left[
 \frac89\log^2x+
 \left(\frac{20}{27}-\frac43c_\infty\right)\log x
 +\mathcal O(1)
 \right],
 \qquad x\to0.
 \label{eq:agamma2_leading_logs}
\end{equation}
In particular, the coefficient $8/9$ of the double logarithm agrees with the second-order expansion of the anomalous dipole falloff $x^\delta$, where
$\delta=1-\frac43\gamma+\frac{20}{27}\gamma^2+\mathcal O(\gamma^3)$.
The additional single-logarithm contribution reflects both the expansion of the exponent and the coupling dependence of the dipole amplitude. This provides a direct asymptotic check of the second-order construction and anticipates the exact-in-$\gamma$ result derived below.
}

At the horizon, $Y_{\mathrm H}g_2$ is regular, and $Y_\infty g_2$ has only an integrable logarithm. Thus $C_2(1)$ is finite, and subtracting it in Eq.~\eqref{eq:agamma2_u2_closed} removes the singular homogeneous contribution. Endpoint values are taken after combining the terms, since individual GPL expressions may diverge separately.

The Wronskian identity~\eqref{eq:agamma2_Wronskian} gives directly $\mathcal D_\rho \widehat v_2=g_2$. Differentiating Eq.~\eqref{eq:agamma2_radial_solution} then yields $w_2'=4Kv_2/r^4$. More explicitly, set $u=x+\gamma \widehat v_1+\gamma^2\widehat v_2$ and $\varepsilon=2(1-e^{-2\gamma})$. The exact dimensionless master equation has the residual
\begin{equation}
 (\mathcal D_\rho+\varepsilon)u+3(1+\rho)x^2
 =\gamma^3\left(4\widehat v_2-4\widehat v_1+\frac83x\right)
 +\mathcal O(\gamma^4).
 \label{eq:agamma2_residual}
\end{equation}
The retained coefficients therefore satisfy the radial field equations analytically through second order in the coupling.

The resulting electric configuration is
\begin{align}
 ds^2={}&-f\,dt^2+\frac{dr^2}{f}
 +r^2\bigl(d\theta^2+\sin^2\theta\,d\varphi^2\bigr)
 -2ar^2\bigl(\omega_{\mathrm s}+\gamma w_1+\gamma^2w_2\bigr)
 \sin^2\theta\,dt\,d\varphi
 +\mathcal O(a^2,a\gamma^3),
\label{eq:agamma2_metric}\\
 A={}&-\frac q r\,dt
 +aq\left(\frac1r+\gamma v_1+\gamma^2v_2\right)
 \sin^2\theta\,d\varphi
 +\mathcal O(a^2,a\gamma^3).
 \label{eq:agamma2_potential}
\end{align}
The logarithmic corrections remain compatible with asymptotic flatness. The nonextremal and asymptotic qualifications discussed in \autoref{sec:agamma} also apply here. The case with nonzero $Q_m$ follows directly by electromagnetic duality, using the general parameters in Eq.~\eqref{eq:defs} and the same radial profiles in Eq.~\eqref{eq:Adyonexact}; no additional radial equations are required.

\subsection{Quadratic rotation at first order in the coupling}
\label{sec:a2gamma}

We next construct the electric solution through $\mathcal O(a^2\gamma)$, using the ansatz~\eqref{eq:metric_a2gamma} and \eqref{eq:potential_a2gamma}. We set $Q_m=0$ in Eq.~\eqref{eq:defs} and use the functions $v_1$ and $w_1$ obtained in \autoref{sec:agamma}. The new fields are the monopole and quadrupole profiles $u_0, u_2, b_0, b_2$, together with the metric function $\mathcal Z$ and the mass-normalisation coefficient $\tau$. The indices $j=0,2$ below label the angular harmonics in Eq.~\eqref{eq:second_order_harmonics}.

The background quantities are those of Eqs.~\eqref{eq:defs}, \eqref{eq:static_horizons}, and~\eqref{eq:reference_dragging}. Regularity conditions for the perturbative coefficients are imposed at $r_+$ order by order, while the horizon of the rotating, mass-normalised geometry is displaced at second order in $a$. Already in the Maxwell limit, this displacement is $r_H=r_+-a^2/(2d)+\mathcal O(a^4)$. All radial expressions refer initially to $r>r_+$, and coefficient values at $r_+$ are obtained by regular limits. As in the preceding constructions, the dependence on $K$ is retained in unexpanded form.

\paragraph{Sourced radial equations.}

At this order, the diagonal metric and electric potential acquire only angular monopole and quadrupole corrections. {These even-parity sectors contain the first non-Maxwell corrections to the electric potential and diagonal geometry associated with rotation. The first-order magnetic and frame-dragging profiles $v_1$ and $w_1$ enter their sources, together with explicit contributions from the ModMax constitutive law. The $j=2$ harmonic here describes an electric and gravitational quadrupolar deformation, rather than a new anomalous magnetic multipole.} Introduce the known source combinations
\begin{align}
 \mathcal F_{\mathrm s} &=-2r^4\omega_{\mathrm s}'w_1' +4K\left[ \left(\frac f{r^2}-\omega_{\mathrm s}\right)v_1' +\frac{w_1}{r^2}\right], \nonumber\\
\mathcal C_{\mathrm s}&=16K\left(\frac1{r^4}-\frac{v_1}{r^3}\right), \qquad \mathcal D_{\mathrm s}=\left(r^2\omega_{\mathrm s}v_1'-w_1\right)'.
\label{eq:a2gamma_source_combinations}
\end{align}
Their harmonic projections are
\begin{align}
 G_0&=\frac{2\mathcal F_{\mathrm s}+\mathcal C_{\mathrm s}}3,
 &G_2&=\frac{-2\mathcal F_{\mathrm s}+2\mathcal C_{\mathrm s}}3,
 \nonumber\\
 E_0&=-\frac23\mathcal D_{\mathrm s},
 &E_2&=-\frac8{r^3}+\frac23\mathcal D_{\mathrm s}
 -\frac4f\left(\omega_{\mathrm s}v_1+\frac{w_1}{r}\right).
\label{eq:a2gamma_harmonic_sources}
\end{align}
For $j=0,2$, with $L_j=j(j+1)$, the coupled gravitational and
electric equations reduce to
\begin{align}
 (\Delta u_j')'
 +\left(\frac{2K}{r^2}-L_j\right)u_j+4Kb_j'&=G_j,
 \label{eq:a2gamma_u_equation}\\
 (r^2b_j')'-\frac{L_j}{f}b_j+u_j'&=E_j.
 \label{eq:a2gamma_b_equation}
\end{align}
These equations determine the harmonic profiles in the intermediate fixed-determinant gauge described in \autoref{sec:ansatz}. The mass adjustment is imposed below.

The system admits an algebraic decoupling. Set
\begin{equation}
 T_j=\left(M-\frac Kr\right)u_j+2Kb_j,\qquad
 N_j=\frac{\Delta}{r}u_j+2(Mr-K)b_j.
 \label{eq:a2gamma_decoupling}
\end{equation}
Then
\begin{equation}
 (\Delta T_j')'-L_jT_j=S_{Tj},\qquad
 \Delta N_j''-L_jN_j=S_{Nj},
\label{eq:a2gamma_decoupled_equations}
\end{equation}
where
\begin{equation}
 S_{Tj}=\left(M-\frac Kr\right)G_j+2KfE_j,\qquad
 S_{Nj}=\frac{\Delta}{r}G_j+2f(Mr-K)E_j.
\label{eq:a2gamma_decoupled_sources}
\end{equation}
The original profiles are recovered from
\begin{equation}
{
 u_j=\frac{(Mr-K)T_j-KN_j}{rd^2},\qquad
 b_j=\frac{-\Delta T_j+(Mr-K)N_j}{2r^2d^2}.
 }
 \label{eq:a2gamma_inverse}
\end{equation}
Although $E_2$ contains a factor $f^{-1}$, the combinations entering Eq.~\eqref{eq:a2gamma_decoupled_sources} have finite horizon limits. We use those combined sources in the expressions below.

\paragraph{Analytic radial quadratures.}

Introduce the radial variable $z=(r-M)/d$ and the Legendre functions
\begin{align}
 \mathsf P_0(r)&=1,
 &\mathsf Q_0(r)&=\frac12\log\frac{z+1}{z-1},
 \nonumber\\
 \mathsf P_2(r)&=\frac12(3z^2-1),
 &\mathsf Q_2(r)&=\mathsf P_2(r)\mathsf Q_0(r)-\frac32z.
 \label{eq:a2gamma_Legendre}
\end{align}
These radial functions are distinct from the angular polynomial $P_2(\cos\theta)$. Their Wronskians obey
\begin{equation}
 \Delta\left(\mathsf P_j\mathsf Q_j'
 -\mathsf P_j'\mathsf Q_j\right)=-d.
\label{eq:a2gamma_Legendre_Wronskian}
\end{equation}
Horizon regularity and asymptotic normalisation select
\begin{equation}
 T_j(r)=-\frac1d\left[
 \mathsf Q_j(r)\int_{r_+}^{r}\mathsf P_j(t)S_{Tj}(t)\,dt
 +\mathsf P_j(r)\int_r^\infty\mathsf Q_j(t)S_{Tj}(t)\,dt
 \right],\qquad j=0,2.
 \label{eq:a2gamma_T_solution}
\end{equation}
For the monopole, define
\begin{equation}
 \tau=-\int_{r_+}^{\infty}S_{T0}(t)\,dt,
 \qquad T_0(r)=\frac{\tau}{r}+o(r^{-1}).
 \label{eq:a2gamma_tau}
\end{equation}
The corresponding second profile is
\begin{equation}
 N_0(r)=\frac{\tau}{M}
 +\int_r^\infty(t-r)\frac{S_{N0}(t)}{\Delta(t)}\,dt.
 \label{eq:a2gamma_N0_solution}
\end{equation}
The constant $\tau/M$ fixes the electric charge by cancelling the $1/r$ contribution to $b_0$. The other homogeneous solution, proportional to $r$, is excluded by asymptotic normalisation. The ratio $S_{N0}/\Delta$ is regular at the horizon.

For the quadrupole, regularity requires a generally nonzero horizon value. Taking the regular limit of its equation gives
\begin{equation}
 n_H\equiv N_2(r_+)
 =-\frac{S_{N2}(r_+)}6
 =\frac43dr_+
 \left[\omega_{\mathrm s}v_1+\frac{w_1}{r}\right]_{r=r_+}.
 \label{eq:a2gamma_N2_horizon}
\end{equation}
To implement this condition explicitly, define
\begin{equation}
 m(r)=\frac{r_+n_H}{r},\qquad
 \widehat S_N=S_{N2}-(\Delta m''-6m),
 \qquad \widehat S_N(r_+)=0,
\label{eq:a2gamma_subtracted_source}
\end{equation}
and introduce the derivative partners
\begin{equation}
 \mathsf P_N=\Delta\mathsf P_2',\qquad
 \mathsf Q_N=\Delta\mathsf Q_2',\qquad
 \mathsf P_N\mathsf Q_N'-\mathsf P_N'\mathsf Q_N=6d.
 \label{eq:a2gamma_N_basis}
\end{equation}
Both solve $\Delta y''-6y=0$. The required solution is
\begin{equation}
 \begin{aligned}
 N_2(r)=m(r)+\frac1{6d}\bigg[
 &\mathsf Q_N(r)\int_{r_+}^{r}
 \mathsf P_N(t)\frac{\widehat S_N(t)}{\Delta(t)}\,dt+\mathsf P_N(r)\int_r^\infty
 \mathsf Q_N(t)\frac{\widehat S_N(t)}{\Delta(t)}\,dt
 \bigg].
 \end{aligned}
 \label{eq:a2gamma_N2_solution}
\end{equation}
This subtraction regularises the horizon limits and introduces no additional parameter. Equations~\eqref{eq:a2gamma_T_solution}, \eqref{eq:a2gamma_N0_solution}, and~\eqref{eq:a2gamma_N2_solution}, followed by the inverse transformation~\eqref{eq:a2gamma_inverse}, determine all four harmonic profiles. Every function appearing under an integral is already known.

\paragraph{Remaining metric function.}

The remaining Einstein constraints determine $\mathcal Z$ algebraically. For compactness, introduce the equatorial combinations
\begin{equation}
 U_e=u_0-\frac12u_2,\qquad B_e=b_0-\frac12b_2,
\label{eq:a2gamma_equatorial_fields}
\end{equation}
and define
\begin{align}
 \mathcal R_e={}&
 \left(\frac1r-\frac{\Delta'}{4\Delta}\right)U_e'
 -\frac{r^4}{2\Delta}\omega_{\mathrm s}'w_1'
 -\frac{2K}{r^4}v_1'
-\frac K\Delta
 \left[
 2B_e'+2\omega_{\mathrm s}v_1'
 -\frac{2w_1}{r^2}+\frac{U_e}{r^2}
 \right],
 \label{eq:a2gamma_Re}\\
 \mathcal C_e={}&
 \frac14\left[
 3\left(\frac2r-\frac{\Delta'}{2\Delta}\right)u_2-U_e'
 \right]
 +2K\left(\frac{v_1}{r^4}-\frac{v_1'}{r^3}\right)
 -\frac K\Delta
 \left[
 3b_2-2\left(\omega_{\mathrm s}v_1+\frac{w_1}{r}\right)
 \right].
 \label{eq:a2gamma_Ce}
\end{align}
Then
\begin{equation}
 \mathcal Z(r)=\frac{\Delta}{(r-M)^2}
 \left[\Delta\mathcal R_e+2(r-M)\mathcal C_e\right].
 \label{eq:a2gamma_Z_nu}
\end{equation}
{The remaining Einstein equations are consistent via a constraint-propagation identity. Defining
\begin{equation}
    \mathcal D(r)=2(T_0'+T_2')-3u_2+2\mathcal Z-\mathcal C_{\mathrm s},  \label{eq:a2gamma_constraint_D}
\end{equation}
where $T_j$ and $\mathcal C_{\mathrm s}$ are defined in Eqs.~\eqref{eq:a2gamma_decoupling} and~\eqref{eq:a2gamma_source_combinations}, respectively, the radial equations~\eqref{eq:a2gamma_u_equation} and~\eqref{eq:a2gamma_b_equation}, together with the algebraic relation~\eqref{eq:a2gamma_Z_nu}, imply
\begin{equation}
    \frac{d}{dr}\left[\Delta(r)(r-M)^2\mathcal D(r)\right]=0. \label{eq:a2gamma_constraint_propagation}
\end{equation}
Consequently, $\mathcal D=C_0/[\Delta(r)(r-M)^2]$, where $C_0$ is an integration constant. Regularity at the nonextremal outer horizon requires $C_0=0$, ensuring that the remaining Einstein constraints are satisfied throughout the exterior. In particular, the Ricci scalar vanishes, consistent with the tracelessness of the ModMax energy-momentum tensor.} The function $\nu$ then follows from Eq.~\eqref{eq:second_order_harmonics}, with no further quadrature. On either symmetry axis, $2\nu=U$, ensuring elementary flatness with the standard $2\pi$ azimuthal period.

\paragraph{Physical normalisation and regularity.}

The monopole profiles have the asymptotic behaviour
\begin{equation}
 u_0=\frac{\tau}{Mr}+o(r^{-1}),\qquad b_0=o(r^{-1}).
\label{eq:a2gamma_monopole_asymptotics}
\end{equation}
Before the mass adjustment included in Eq.~\eqref{eq:metric_a2gamma}, the resulting ADM mass is
\begin{equation}
M_{\mathrm{ADM}}=M+\frac{\gamma a^2\tau}{2M}.
\label{eq:a2gamma_intermediate_mass}
\end{equation}
Subtracting the homogeneous Reissner--Nordstr\"om mass variation $\delta M=\gamma a^2\tau/(2M)$ produces precisely
\begin{equation}
 \delta(ds^2)_{\mathrm{mass}}
 =-\frac{\gamma a^2\tau}{Mr}\,dt^2
 -\frac{\gamma a^2\tau}{Mrf^2}\,dr^2.
\label{eq:a2gamma_mass_adjustment}
\end{equation}
These are the two counterterms already displayed in the ansatz. They keep $M$ equal to the physical ADM mass and leave the electromagnetic potential unchanged at the retained order. Their effect on $J=aM$ starts at $\mathcal O(a^3\gamma)$. Thus $\tau$ is fixed by the sources and introduces no independent parameter.

The quadrupole horizon condition~\eqref{eq:a2gamma_N2_horizon} implies
\begin{equation}
 b_2(r_+)=\frac23
 \left[\omega_{\mathrm s}v_1+\frac{w_1}{r}\right]_{r=r_+},
 \qquad
 \mathcal Z(r_+)=-\frac32u_2(r_+).
\label{eq:a2gamma_horizon_checks}
\end{equation}
The first relation cancels the quadrupolar contribution to $A_t+\Omega_HA_\varphi$ on the horizon. The second makes $U/2+\nu=u_0+u_2$ independent of angle there, as required by the surface-gravity condition in the intermediate fixed-determinant gauge. The subsequent mass adjustment need not preserve that gauge and entails a perturbative horizon displacement. In particular, the $f^{-2}$ term in Eq.~\eqref{eq:a2gamma_mass_adjustment} is a coordinate artifact of this expansion; we assess horizon regularity in a regular ingoing chart.

Substituting the profiles above into Eqs.~\eqref{eq:metric_a2gamma} and~\eqref{eq:potential_a2gamma} completes the electric solution through $\mathcal O(a^2\gamma)$, with rotational remainders $\mathcal O(a^3,a\gamma^2)$. The displayed Green functions solve the radial equations, and the remaining constraints fix $\mathcal Z$ as above. The resulting geometry is asymptotically flat with the prescribed mass, electric charge, and angular momentum. All radial functions are specified by explicit, boundary-normalised analytic quadratures with elementary kernels and known logarithmic and dilogarithmic sources. The boundary conditions fix every integration constant. {The quadratic-order solution thus makes explicit how the nonlinear response responsible for the modified rotation-induced magnetic dipole feeds back into the electrostatic potential and diagonal geometry. These monopole and quadrupole corrections are determined by the lower-order rotational profiles and the field equations, rather than by independent electromagnetic multipole parameters.} The construction assumes a nonextremal charged seed; the extremal limit requires a separate analysis.

\paragraph{Outer horizon and angular velocity.}

We now describe the main geometrical properties of the mass-normalised solution. For brevity, write $h=r_+$ using Eq.~\eqref{eq:static_horizons}, and assume $a^2\ll d^2$ within the nonextremal branch. The symbol $\doteq$ below denotes equality through the retained orders: the omitted even-in-$a$ metric terms are $\mathcal O(a^4,a^2\gamma^2)$, while the omitted frame-dragging terms are $\mathcal O(a^3,a\gamma^2)$. The time coordinate is nonperiodic and normalised at infinity, and
$\varphi$ has period $2\pi$.

Introduce
\begin{align}
\Delta_{\mathrm{rot}}&=\Delta+a^2\left(1+\frac{\gamma\tau r}{M}\right),
&\Sigma_\nu&=r^2+a^2\cos^2\theta+2\gamma a^2r^2\nu,
\label{eq:a2gamma_geometry_denominators}\\
 \mathcal F(r,\theta)&= 1+\frac{a^2}{r^2}\left[1+(1-f)\sin^2\theta\right] +\gamma a^2U,
&\varpi(r,\theta)&=\frac{a(\omega_{\mathrm s}+\gamma w_1)}{\mathcal F}. \label{eq:a2gamma_geometry_factors}
\end{align}
A useful representative of the perturbative metric is
\begin{equation}
 d\widehat s^2=
 -\frac{\Delta_{\mathrm{rot}}}{r^2\mathcal F}\,dt^2
 +r^2\mathcal F\sin^2\theta\,(d\varphi-\varpi\,dt)^2
 +\Sigma_\nu\left(\frac{dr^2}{\Delta_{\mathrm{rot}}}+d\theta^2\right).
\label{eq:a2gamma_geometry_representative}
\end{equation}
Its expansion reproduces Eq.~\eqref{eq:metric_a2gamma}; the field equations do not determine the higher-order terms implicit in this form. In particular, $g^{rr}\doteq\Delta_{\mathrm{rot}}/\Sigma_\nu$ and $g_{tt}g_{\varphi\varphi}-g_{t\varphi}^2 \doteq-\Delta_{\mathrm{rot}}\sin^2\theta$. The apparent poles at $f=0$ in the expanded metric therefore encode a displaced radial denominator.

The horizon value of $w_1$ and the integral in Eq.~\eqref{eq:a2gamma_tau} simplify to
\begin{equation}
 w_1(h)=-\frac{8K}{9Mh^3},
 \qquad
 \tau=-\frac{8K(M+2d)}{9Mh^2}.
\label{eq:a2gamma_geometry_horizon_data}
\end{equation}
The outer zero of $\Delta_{\mathrm{rot}}$, expanded consistently, and its angular velocity are consequently
\begin{align}
 r_H&=h-\frac{a^2}{2d}
 \left[1-\frac{8\gamma K(M+2d)}{9M^2h}\right]
 +\mathcal O(a^4,a^2\gamma^2),
\label{eq:a2gamma_geometry_horizon}\\
 \Omega_H&=-\left.\frac{g_{t\varphi}}{g_{\varphi\varphi}}\right|_{r_H}
 =\frac{a}{h^2}\left(1-\frac{8\gamma K}{9Mh}\right)
 +\mathcal O(a^3,a\gamma^2).
\label{eq:a2gamma_geometry_angular_velocity}
\end{align}
Both are independent of $\theta$ at the stated order. {In particular, the correction to the horizon angular velocity is controlled by the frame-dragging profile $w_1$ obtained at first order in rotation, whereas the horizon displacement also depends on the quadratic-order mass adjustment through $\tau$.} The displacement $r_H-h=\mathcal O(a^2)$ affects $\Omega_H$ only at $\mathcal O(a^3)$, which lies beyond the present solution. The associated Killing generator satisfies
\begin{equation}
\chi=\partial_t+\Omega_H\partial_\varphi,  \qquad \left.\chi^\mu\chi_\mu\right|_{r_H} =\left.(g_{tt}+2\Omega_Hg_{t\varphi} +\Omega_H^2g_{\varphi\varphi})\right|_{r_H}  =\mathcal O(a^4,a^2\gamma^2).
\label{eq:a2gamma_geometry_null_generator}
\end{equation}
Thus the displaced surface is the outer Killing horizon to the retained order, using the regular local continuation of the radial profiles. This exterior construction does not establish a regular inner horizon or an extremal limit.

\paragraph{Stationary-limit surface and ergoregion.}

The outer stationary-limit surface is determined by $g_{tt}=0$. Its location is
\begin{equation}
 r_E(\theta)=h-\frac{a^2\cos^2\theta}{2d}
 +\frac{4\gamma a^2K}{9M^2hd}
 \left(2d-M+2M\cos^2\theta\right)
 +\mathcal O(a^4,a^2\gamma^2).
\label{eq:a2gamma_geometry_ergosurface}
\end{equation}
The coordinate separation from the outer horizon is
\begin{equation}
 r_E(\theta)-r_H
 =\frac{a^2\sin^2\theta}{2d}
 \left(1-\frac{16\gamma K}{9Mh}\right)
 +\mathcal O(a^4,a^2\gamma^2).
\label{eq:a2gamma_geometry_ergoregion}
\end{equation}
The surfaces meet on both halves of the symmetry axis, and their separation is maximal at the equator. In the controlled perturbative regime, the intervening region has $g_{tt}>0$. At fixed $K$, the explicit nonlinear correction reduces this coordinate separation; it does not necessarily move the stationary-limit surface inward, since its equatorial correction is proportional to $2d-M$. The separation is not an invariant thickness, and a zero of its truncated prefactor cannot be used to infer disappearance of the ergoregion.

These expressions use the same unexpanded static parameter $K$ as the solution. A strict Taylor expansion at fixed $M,Q_e$, and $a$ follows by expanding every occurrence of $K$, $d$, and $h$ consistently; it additionally requires $M^2>Q_e^2$ and $|\gamma|Q_e^2\ll M^2-Q_e^2$. No separate expansion of the geometrical equations is needed.

\paragraph{Axis regularity and exterior causality.}

The squared norm of the axial Killing vector is $X=g_{\varphi\varphi}\doteq r^2\mathcal F\sin^2\theta$. Define $U_{\mathrm{ax}}=u_0+u_2$ and use the equatorial combination $U_e$ from Eq.~\eqref{eq:a2gamma_equatorial_fields}. The angular dependence obeys
\begin{equation}
 \begin{aligned}
 \mathcal F&=\mathcal F_{\mathrm{ax}}\cos^2\theta
 +\mathcal F_{\mathrm{eq}}\sin^2\theta,\\
 \mathcal F_{\mathrm{ax}}&=1+\frac{a^2}{r^2}
 +\gamma a^2U_{\mathrm{ax}},
 \qquad
 \mathcal F_{\mathrm{eq}}=1+\frac{a^2}{r^2}(2-f)
 +\gamma a^2U_e.
\end{aligned} \label{eq:a2gamma_geometry_axial_factors}
\end{equation}
Near either axis, with $\epsilon=\theta$ or $\pi-\theta$, the axis value $2\nu|_{\mathrm{axis}}=U_{\mathrm{ax}}$ gives $X=C(r)\epsilon^2+\mathcal O(\epsilon^4)$ and $g_{\theta\theta}=C(r)+\mathcal O(\epsilon^2)$, where $C=r^2\mathcal F_{\mathrm{ax}}$. Hence, for $C>0$,
\begin{equation}
\lim_{\epsilon\to0}\frac{\nabla_\mu X\nabla^\mu X}{4X}
 \doteq1.
\label{eq:a2gamma_geometry_elementary_flatness}
\end{equation}
Both halves of the axis are therefore elementary flat with period $2\pi$. Moreover, $g_{t\varphi}$ and $A_\varphi$ vanish as $\epsilon^2$, so the electric configuration has neither an axial time twist nor an electromagnetic string.

Positivity of $\mathcal F_{\mathrm{ax}}$ and $\mathcal F_{\mathrm{eq}}$ throughout the exterior ensures $X>0$ away from the axis and excludes timelike azimuthal circles. A uniform sufficient condition is
\begin{equation}
 \gamma a^2\mathcal U_-<1,
 \qquad
 \mathcal U_-=
 \sup_{r\geq r_0}\max\{0,-U_{\mathrm{ax}}(r),-U_e(r)\},
\label{eq:a2gamma_geometry_positivity}
\end{equation}
where $M<r_0<h$ lies in the regular local continuation and $r_H>r_0$. Regularity and asymptotic decay make $\mathcal U_-$ finite, while $1-f>0$ on this interval. The inequality must hold with a margin relative to the neglected terms; it does not assume that the nonlinear correction itself is positive.

For the representative~\eqref{eq:a2gamma_geometry_representative}, the conditions $\mathcal F>0$, $\Sigma_\nu>0$, and $\Delta_{\mathrm{rot}}>0$ outside its outer root give
\begin{equation}
 \widehat g^{tt}
 =-\frac{r^2\mathcal F}{\Delta_{\mathrm{rot}}}<0.
\label{eq:a2gamma_geometry_temporal_function}
\end{equation}
The globally defined, nonperiodic coordinate $t$ is therefore a temporal function, establishing stable causality and excluding all closed causal curves in this exterior~\cite{Minguzzi:2010juk}.
This remains true within the ergoregion, where $g_{tt}>0$. The causal statement applies to this Lorentzian representative of the retained perturbative coefficients. It does not establish causality for an unconstructed all-orders solution or inside the horizon; an exact completion requires corresponding control of the omitted terms, particularly near the displaced outer surface.

\subsection{Linear rotation at arbitrary coupling}
\label{sec:finite}

We finally retain arbitrary finite $\gamma\geq0$ while truncating only at first order in $a$. The radial equations \eqref{eq:agamma_coupled} remain valid without a coupling expansion. We use the electric specialisation of Eq.~\eqref{eq:defs}; only the screened nonextremality condition $0<K<M^2$ is required. Eliminating $\omega'$ using Eq.~\eqref{eq:agamma_spin_normalisation} gives
\begin{equation}
 (fV')'-\left(\frac{\lambda}{r^2}+\frac{4K}{r^4}\right)V
 =-\frac{6M}{r^4},\qquad
 \lambda=2e^{-2\gamma}.
 \label{eq:finite_master}
\end{equation}
The boundary conditions are regularity at the outer horizon and the exclusion of the growing electromagnetic mode at infinity.

Using $x$ and $\rho$ from Eq.~\eqref{eq:agamma2_variables}, define
\begin{equation}
 \mathcal U(x)=r_+V(r),\qquad
\delta=\frac{\sqrt{1+8e^{-2\gamma}}-1}{2}.
 \label{eq:finite_variables}
\end{equation}
Here $\lambda=\delta(\delta+1)$ and $0<\delta\leq1$ for finite $\gamma\geq0$. With $p(x)$ from Eq.~\eqref{eq:agamma2_operator}, Eq.~\eqref{eq:finite_master} becomes
\begin{equation}
\partial_x\!\left[p(x)\partial_x\mathcal U\right]
 -(\lambda+4\rho x^2)\mathcal U=-3(1+\rho)x^2.
 \label{eq:finite_dimensionless}
\end{equation}

{The exponent $\delta$ has a direct electromagnetic interpretation. At large radius, the leading homogeneous part of Eq.~\eqref{eq:finite_master} reduces to $V''-\lambda V/r^2=0$, which is precisely the equation governing a weak magnetic dipole perturbation around an electric Coulomb background in flat-space ModMax electrodynamics, as discussed in the Appendix. The fractional exponent is therefore determined by the electromagnetic constitutive response, rather than by gravitational curvature. The black hole geometry modifies the radial profile and fixes its amplitude through the boundary conditions.}

\paragraph{Exterior Frobenius representation.}

The homogeneous exponents at $x=0$ are $\delta$ and $-1-\delta$. Excluding the growing mode gives
\begin{equation}
 V(r)=\frac1{r_+}\left[
 C_\gamma x^\delta\sum_{n=0}^{\infty}c_nx^n
 +x^2\sum_{n=0}^{\infty}d_nx^n
 \right],
 \label{eq:finite_V_series}
\end{equation}
where $C_\gamma$ also depends on $\rho$. The coefficients are fixed recursively by
\begin{equation}
 c_{-1}=d_{-1}=0,\qquad c_0=1,\qquad
 d_0=-\frac{3(1+\rho)}{6-\lambda},
 \label{eq:finite_series_seeds}
\end{equation}
and, for $n\geq1$,
\begin{equation}
 e_n=
 \frac{(1+\rho)(p+n-1)(p+n+1)e_{n-1}
 -\rho(p+n+2)(p+n-3)e_{n-2}}
 {(p+n)(p+n+1)-\lambda}.
 \label{eq:finite_recurrence}
\end{equation}
The choices $(p,e_n)=(\delta,c_n)$ and $(2,d_n)$ give the homogeneous and particular series, respectively. For the former, the denominator is $n(n+2\delta+1)>0$; for the latter, it is $(n+2)(n+3)-\lambda>0$. The initial particular coefficient is also regular because $6-\lambda\geq4$.

The differential equation has finite singular points at $x=0,1,\rho^{-1}$. Since $\rho^{-1}>1$, both power series in Eq.~\eqref{eq:finite_V_series} converge for $0\leq x<1$. They therefore provide a representation at every exterior radius $r>r_+$. Term-by-term differentiation and integration are valid on compact subsets of this open exterior.

\paragraph{Horizon connection coefficient.}

At $x=1$, the homogeneous indicial exponent is repeated. The two terms in Eq.~\eqref{eq:finite_V_series} generally contain logarithms. Let $A_c$ and $A_d$ denote their respective coefficients of $\log(1-x)$ before multiplication by $C_\gamma$. Horizon regularity requires
\begin{equation}
 C_\gamma A_c+A_d=0,\qquad
 C_\gamma=-\frac{A_d}{A_c}
 =-\lim_{n\to\infty}\frac{d_n}{c_n}.
 \label{eq:finite_connection}
\end{equation}
The last equality follows from $c_n\sim-A_c/n$ and $d_n\sim-A_d/n$. Multiplication by $x^\delta$ or $x^2$ does not change the leading logarithmic amplitude at $x=1$. Equation~\eqref{eq:finite_connection} defines the connection coefficient exactly in terms of the recurrence.

The cancellation is unique. If $\psi$ were a homogeneous solution of Eq.~\eqref{eq:finite_master}, regular at $r_+$ and decaying at infinity, integration by parts would give
\begin{equation}
 \int_{r_+}^{\infty}
 \left[
 f(\psi')^2+
 \left(\frac{\lambda}{r^2}+\frac{4K}{r^4}\right)\psi^2
 \right]dr=0.
 \label{eq:finite_uniqueness}
\end{equation}
The boundary terms vanish, and the integrand is non-negative, so $\psi=0$. In particular, the nonzero decaying homogeneous solution cannot be regular at the horizon, implying $A_c\neq0$. Thus $C_\gamma$ is fixed rather than an additional parameter.

{This has a useful physical consequence: within the present family, the amplitude of the anomalous magnetic dipole is not an independent parameter. Its value is selected by horizon regularity once the mass, charge, angular momentum, and ModMax coupling are specified. The constitutive law fixes the fractional exponent, while the connection coefficient encodes the response of the black hole geometry.}

The two series should be combined before taking the horizon limit; their individual sums generally diverge there. After imposing Eq.~\eqref{eq:finite_connection}, the regular continuation satisfies
\begin{equation}
 (1-\rho)\,\partial_x\mathcal U(1)
 +(\lambda+4\rho)\mathcal U(1)=3(1+\rho).
\label{eq:finite_horizon_condition}
\end{equation}
Higher horizon coefficients follow recursively from Eq.~\eqref{eq:finite_dimensionless}.

\paragraph{Metric reconstruction and asymptotics.}

The integrated Einstein equation fixes the frame-dragging profile:
\begin{equation}
 \begin{aligned}
 \omega(r)=\frac1{r_+^2}\bigg[
 &(1+\rho)x^3
 -4\rho C_\gamma\sum_{n=0}^{\infty}
 \frac{c_nx^{n+\delta+3}}{n+\delta+3}
 -4\rho\sum_{n=0}^{\infty}
 \frac{d_nx^{n+5}}{n+5}
 \bigg].
 \end{aligned}
 \label{eq:finite_omega_series}
\end{equation}
Equations~\eqref{eq:finite_V_series} and \eqref{eq:finite_omega_series}, inserted into Eqs.~\eqref{eq:metric} and~\eqref{eq:Adyonexact}, determine the metric and electromagnetic field through first order in rotation. The dyonic extension uses the general parameters of Eq.~\eqref{eq:defs} and the same radial construction, without additional radial equations or connection parameters.

The inhomogeneous equation and the boundary conditions imply $V>0$ for $M>0$. Since the particular infinity coefficient $d_0$ is negative, this also implies $C_\gamma>0$. Writing $b=C_\gamma r_+^{\delta-1}$, the leading falloffs are
\begin{equation}
 V\sim b\,r^{-\delta},\qquad
 \omega=\frac{2M}{r^3}
 -\frac{4Kb}{3+\delta}r^{-3-\delta}
 +o(r^{-3-\delta}),
 \qquad
 \omega-fV'\sim\delta b\,r^{-1-\delta}.
 \label{eq:finite_asymptotics}
\end{equation}
The induced magnetic potential consequently decays as $r^{-\delta}$, and its dyonic electric-dipole counterpart as $r^{-1-\delta}$. The corresponding orthonormal field corrections decay as $r^{-2-\delta}$. In particular,
\begin{equation}
 g_{t\varphi}=-\frac{2J}{r}\sin^2\theta
 +\mathcal O(r^{-1-\delta}),\qquad J=aM.
\label{eq:finite_metric_asymptotics}
\end{equation}
Because $\delta>0$ at every finite coupling, these electromagnetic tails remain compatible with asymptotic flatness and the prescribed mass and angular momentum.

{
More explicitly, for the electrically charged solution, the leading asymptotic magnetic field follows from $A_\varphi=aqV(r)\sin^2\theta$. Its orthonormal components are
\begin{equation}
 B^{\hat r}\sim
 \frac{2aqb}{r^{2+\delta}}\cos\theta,
 \qquad
 B^{\hat\theta}\sim
 \frac{aq\delta b}{r^{2+\delta}}\sin\theta.
 \label{eq:finite_magnetic_components}
\end{equation}
The field retains the angular structure of a magnetic dipole, but its radial falloff and the relative amplitudes of its components differ from the Maxwell values whenever $\gamma>0$. In particular, extracting a magnetic moment through the conventional Maxwell dipole formulas would give quantities proportional to $r^{1-\delta}$, which neither agree between the two components nor approach a finite constant. Thus, the anomalous magnetic response cannot be described by a conventional constant magnetic dipole moment or gyromagnetic ratio. Instead, its asymptotic amplitude is characterised by $aqb=aqC_\gamma r_+^{\delta-1}$, fixed by the black hole boundary conditions. A more detailed flat-spacetime interpretation is given in the Appendix.
}

At $\gamma=0$, one has $\delta=1$ and $\lambda=2$. The recurrence gives $d_n=-c_{n+1}$, while horizon regularity fixes $C_0=1$. The two series then reduce to $\mathcal U=x$, recovering $V=1/r$ and $\omega=2M/r^3-K/r^4$. For small coupling,
\begin{equation}
 \delta=1-\frac43\gamma+\frac{20}{27}\gamma^2
 +\mathcal O(\gamma^3),
 \qquad
 x^\delta=x\left[
 1-\frac43\gamma\log x
+\gamma^2\left(\frac89\log^2x+\frac{20}{27}\log x\right)
 +\mathcal O(\gamma^3)
 \right].
 \label{eq:finite_small_gamma}
\end{equation}
This expansion explains the logarithmic asymptotic terms in the preceding coupling expansions.

{More precisely, the logarithmic coefficients arise from expanding both the fractional power and its coupling-dependent amplitude $C_\gamma$. The leading double logarithm at order $\gamma^2$ has the universal coefficient $8/9$, in agreement with the asymptotic result of \autoref{sec:agamma2}. At the same time, the single logarithms also depend on the lower-order expansion of $C_\gamma$. The small-coupling expansion is therefore nonuniform at spatial infinity: at any fixed nonzero $\gamma$, sufficiently large radii invalidate an expansion in powers of $\gamma\log(r/r_+)$. The finite-coupling solution instead retains the fractional power and provides the appropriate asymptotic description.}

Expanding the complete radial equation and its boundary conditions reproduces the perturbative problems of Secs.~\ref{sec:agamma} and~\ref{sec:agamma2}. The series construction is exact in $\gamma$ at first order in $a$; its convergence statement concerns the nonextremal exterior. The extremal and finite-rotation problems require separate treatment.

\section{Final remarks}
\label{sec:final_remarks}

We have constructed analytic rotational deformations of nonextremal Einstein--ModMax black holes by retaining independent frame-dragging and rotation-induced magnetic profiles. This freedom accommodates the nonlinear constitutive response that obstructs the screened Kerr--Newman ansatz. The coupled field equations are solved at the retained perturbative orders, with horizon regularity and asymptotic normalisation selecting the solutions for fixed physical parameters.

At first order in rotation, the coupling expansion yields finite closed expressions up to $\mathcal O(a\gamma)$ in logarithms and dilogarithms, and up to $\mathcal O(a\gamma^2)$ in Goncharov multiple polylogarithms. Electromagnetic duality supplies the corresponding dyonic configurations without additional radial equations. In the electric sector, the $\mathcal O(a^2\gamma)$ extension resolves the monopole and quadrupole corrections to the geometry and electric potential in explicit analytic radial quadratures. These complementary constructions describe successive departures from Kerr--Newman while keeping the distinction between finite closed expressions and boundary-normalised quadratures explicit. {In particular, the quadratic-order solution shows how the nonlinear electromagnetic response responsible for the modified rotation-induced magnetic field also affects the electrostatic potential and diagonal geometry.}

Keeping $\gamma$ finite at first order in rotation gives a convergent exterior series representation with recursively determined coefficients. Horizon regularity fixes the connection constant that relates the asymptotic and horizon solutions. {A particularly significant feature of this solution is the anomalous asymptotic behaviour of the rotation-induced magnetic dipole, whose field decays as $r^{-2-\delta}$, with $\delta=(\sqrt{1+8e^{-2\gamma}}-1)/2$. For nonzero $\gamma$, this differs from the standard Maxwell dipole falloff $r^{-3}$ and cannot be characterised by a conventional constant magnetic dipole moment or gyromagnetic ratio. As we have shown, the anomalous exponent originates in the anisotropic constitutive response of ModMax around an electric Coulomb background, a phenomenon already present in flat spacetime. The electromagnetic theory determines the fractional exponent, while the black hole boundary conditions fix the amplitude of the corresponding mode. This provides a physical explanation for the incompatibility of Kerr--Newman-type magnetic profiles with the rotating Einstein--ModMax equations.} {The small-coupling expansion of the fractional power reproduces the logarithmic terms found in the perturbative profiles, revealing their common physical origin. This expansion is nonuniform at spatial infinity, whereas the finite-coupling solution retains the correct fractional-power decay.} {Despite the modified electromagnetic asymptotics, the geometry remains asymptotically flat, with well-defined mass and angular momentum.}

The construction applies to charged, nonextremal backgrounds and does not determine the finite-rotation geometry. Extending it to the extremal regime requires a separate treatment of the horizon boundary conditions. Higher orders in rotation and the full stationary, axisymmetric boundary-value problem remain natural directions for further work. {In particular, the anomalous magnetic asymptotics provide an additional constraint on possible exact rotating solutions, whose electromagnetic sector must accommodate radial behaviour beyond the conventional Kerr--Newman multipole structure.} The independent metric and electromagnetic profiles identified here provide the starting point for these extensions.

\acknowledgments
J.B. and A.C. appreciate discussions with Omkar Shetye, and thank Jagiellonian University, where this work was initiated. J.B. acknowledges the support of ANID through grant FOVI250117 and the hospitality of the Institute of Mathematics, Czech Academy of Sciences (RVO 67985840). A.C. is partially supported by FONDECYT grant 1250318. C.H. is supported by CIDMA (\url{https://ror.org/05pm2mw36}) under the Portuguese Foundation for Science and Technology (FCT, \url{https://ror.org/ 00snfqn58}), Grants UID/04106/2025 (\url{https://doi.org/10.54499/UID/ 04106/2025}) and UID/PRR/04106/2025 (\url{https://doi.org/10.54499/UID/PRR/ 04106/2025}), as well as the Horizon Europe staff exchange (SE) programme HORIZON-MSCA2021-SE-01 Grant No.\ NewFunFiCO-101086251.

\appendix

\section{Electromagnetic duality and the induced electric dipole}
\label{app:duality_dipole}

This appendix derives the combination $\omega-fV'$ appearing in the dyonic potential~\eqref{eq:Adyonexact}. The calculation retains arbitrary $\gamma$ and is carried out to first order in $a$. Unless indicated otherwise, all expressions below are understood modulo $\mathcal O(a^2)$.

Consider an electric representative of physical charge $Q>0$, with
\begin{equation}
 q_0=e^{-\gamma}Q,\qquad K=e^{-\gamma}Q^2,\qquad
 f=1-\frac{2M}{r}+\frac{K}{r^2}.
\label{eq:app_dipole_seed_parameters}
\end{equation}
The metric and potential are
\begin{align}
 ds^2={}&-f\,dt^2+\frac{dr^2}{f}
+r^2\bigl(d\theta^2+\sin^2\theta\,d\varphi^2\bigr)-2ar^2\omega\sin^2\theta\,dt\,d\varphi,
\label{eq:app_dipole_metric}\\
 A_{\mathrm e}={}&-\frac{q_0}{r}\,dt
+aq_0V(r)\sin^2\theta\,d\varphi.
\label{eq:app_dipole_seed_potential}
\end{align}
The independent nonzero components of $F=dA_{\mathrm e}$ are
\begin{equation}
 F_{tr}=-\frac{q_0}{r^2},\qquad
F_{r\varphi}=aq_0V'\sin^2\theta,\qquad F_{\theta\varphi}=2aq_0V\sin\theta\cos\theta.
\label{eq:app_dipole_F}
\end{equation}
The electromagnetic invariants consequently satisfy
\begin{equation}
 \mathcal S=-\frac{q_0^2}{2r^4}+\mathcal O(a^2),
 \qquad
 \mathcal P=\frac{2aq_0^2V}{r^4}\cos\theta+\mathcal O(a^2).
\label{eq:app_dipole_invariants}
\end{equation}
Since the background lies in the electric branch $\mathcal S<0$, the constitutive relation reduces to
\begin{equation}
 H_{\mu\nu}
 =e^\gamma F_{\mu\nu}
-4aV\cos\theta\,\sinh\gamma\,({}^{\star}F)_{\mu\nu}
 +\mathcal O(a^2).
\label{eq:app_dipole_constitutive}
\end{equation}
Its nonzero independent components at the retained order are
\begin{equation}
 H_{tr}=-\frac{Q}{r^2},\qquad
H_{r\varphi}=aQV'\sin^2\theta,\qquad
H_{\theta\varphi}=2aQe^{-2\gamma}V\sin\theta\cos\theta.
\label{eq:app_dipole_H}
\end{equation}
In particular, using $({}^{\star}F)_{\theta\varphi}=q_0\sin\theta$,
\begin{align}
 H_{\theta\varphi}
&=2aq_0V\sin\theta\cos\theta
 \bigl(e^\gamma-2\sinh\gamma\bigr)
 \nonumber\\
 &=2aq_0e^{-\gamma}V\sin\theta\cos\theta.
\label{eq:app_dipole_angular_response}
\end{align}
This component therefore has a different constitutive factor from $H_{tr}$ and $H_{r\varphi}$.

{The unequal constitutive factors in the magnetic components are the covariant manifestation of the anisotropic magnetic response around an electric Coulomb background. In an orthonormal frame, the radial and tangential magnetic perturbations obey $H_{\hat r}=e^{-\gamma}B_{\hat r}$ and $H_{\hat\theta}=e^\gamma B_{\hat\theta}$, respectively. This anisotropy is responsible for the anomalous magnetic dipole falloff discussed in the main text and derived independently in flat spacetime.}

The inverse metric contains
\begin{equation}
 g^{tt}=-\frac1f,\qquad g^{rr}=f,\qquad
g^{\varphi\varphi}=\frac1{r^2\sin^2\theta},\qquad
 g^{t\varphi}=-\frac{a\omega}{f}.
\label{eq:app_dipole_inverse_metric}
\end{equation}
Raising the indices of the excitation requires the off-diagonal term:
\begin{align}
 H^{r\varphi}
 &=g^{rr}\bigl(
g^{\varphi\varphi}H_{r\varphi}
 +g^{\varphi t}H_{rt}\bigr)
 \nonumber\\
&=f\left[\frac{aQV'\sin^2\theta}{r^2\sin^2\theta}
 -\frac{a\omega}{f}\frac{Q}{r^2}\right]
 =\frac{aQ}{r^2}(fV'-\omega).
\label{eq:app_dipole_raised_H}
\end{align}
With the orientation
$\epsilon_{tr\theta\varphi}=+\sqrt{-g}=r^2\sin\theta$
at this order, it follows that
\begin{equation}
({}^{\star}H)_{t\theta}
={\epsilon_{t\theta r\varphi}}H^{r\varphi}
 =-r^2\sin\theta\,H^{r\varphi}
 =aQ(\omega-fV')\sin\theta.
\label{eq:app_dipole_dual_ttheta}
\end{equation}
Thus, the contribution proportional to $\omega$ originates from $g^{t\varphi}H_{rt}$, while the contribution proportional to $fV'$ comes from the radial magnetic component.

Define
\begin{equation}
 W(r)=\omega(r)-f(r)V'(r).
\label{eq:app_dipole_W}
\end{equation}
The remaining nonzero independent components of the dual excitation are
\begin{equation}
 ({}^{\star}H)_{tr}
=\frac{2aQe^{-2\gamma}}{r^2}V\cos\theta,
 \qquad
({}^{\star}H)_{\theta\varphi}=Q\sin\theta.
\label{eq:app_dipole_dual_components}
\end{equation}
On a solution of the electromagnetic equation, $d({}^{\star}H)=0$, a local potential $\widetilde A$ exists with $d\widetilde A={}^{\star}H$. Its $t\theta$ component gives
\begin{equation}
 -\partial_\theta\widetilde A_t=aQW\sin\theta,\qquad
 \widetilde A_t=aQW(r)\cos\theta+\chi(r).
\label{eq:app_dipole_angular_integration}
\end{equation}
{The $\cos\theta=P_1(\cos\theta)$ dependence identifies the dipolar sector of the dual potential. Under the duality rotation constructed below, this term generates the electric dipole associated with a nonzero magnetic monopole charge.}
The $\theta\varphi$ component fixes $\widetilde A_\varphi=-Q\cos\theta$, up to a gauge-patch constant. Matching the $tr$ component then gives
\begin{equation}
 W'=-\frac{2e^{-2\gamma}}{r^2}V,\qquad \chi'=0.
\label{eq:app_dipole_radial_consistency}
\end{equation}
{A gauge transformation can remove the constant $\chi$.}
The first relation is precisely the radial electromagnetic equation,
\begin{equation}
 (fV')'-\frac{2e^{-2\gamma}}{r^2}V=\omega'.
\label{eq:app_dipole_Maxwell}
\end{equation}
A local dual potential is, therefore,
\begin{equation}
 \widetilde A=aQ(\omega-fV')\cos\theta\,dt-Q\cos\theta\,d\varphi.
\label{eq:app_dipole_dual_potential}
\end{equation}
Indeed,
\begin{equation}
 d\widetilde A =-aQW'\cos\theta\,dt\wedge dr
 +aQW\sin\theta\,dt\wedge d\theta
 +Q\sin\theta\,d\theta\wedge d\varphi,
\label{eq:app_dipole_dual_check}
\end{equation}
which agrees with the dual excitation upon using Eq.~\eqref{eq:app_dipole_radial_consistency}.

The ModMax duality rotation acts on the closed two-forms as
\begin{equation}
 \begin{pmatrix}
 F_\alpha\\ {}^{\star}H_\alpha
 \end{pmatrix}
 =
 \begin{pmatrix}
\cos\alpha&\sin\alpha\\-\sin\alpha&\cos\alpha
 \end{pmatrix}
 \begin{pmatrix}
 F\\ {}^{\star}H
 \end{pmatrix}.
\label{eq:app_dipole_duality}
\end{equation}
Choose
\begin{equation}
Q=\sqrt{Q_e^2+Q_m^2},\qquad
\cos\alpha=\frac{Q_e}{Q},\qquad
\sin\alpha=\frac{Q_m}{Q}.
\label{eq:app_dipole_charge_rotation}
\end{equation}
The flux definitions then give electric and magnetic charges $Q_e$ and $Q_m$, respectively. Since the stress tensor is invariant, the metric is unchanged, with $K=e^{-\gamma}(Q_e^2+Q_m^2)$. A corresponding local gauge potential is
\begin{equation}
A_\alpha=\cos\alpha\,A_{\mathrm e}+\sin\alpha\,\widetilde A,
\label{eq:app_dipole_potential_rotation}
\end{equation}
or explicitly,
\begin{equation}
 \begin{aligned}
 A_\alpha={}&
 \left[-\frac{e^{-\gamma}Q_e}{r}
 +aQ_m(\omega-fV')\cos\theta\right]dt\\
 &+\left[-Q_m\cos\theta
 +ae^{-\gamma}Q_eV\sin^2\theta\right]d\varphi
 +\mathcal O(a^2).
 \end{aligned}
\label{eq:app_dipole_dyonic_potential}
\end{equation}
This recovers Eq.~\eqref{eq:Adyonexact}. For nonzero magnetic charge, the potential is understood in the patches $A_N=A_\alpha+Q_m\,d\varphi$ and $A_S=A_\alpha-Q_m\,d\varphi$.

{The same anomalous asymptotic scaling also appears in the induced electric dipole. For the arbitrary-coupling solution of \autoref{sec:finite}, one has $V\sim b\,r^{-\delta}$ and $\omega-fV'\sim\delta b\,r^{-1-\delta}$, with $b=C_\gamma r_+^{\delta-1}$. Hence the dipolar contribution to the dyonic electric potential behaves as}
\begin{equation}
 {
 A_t^{\mathrm{(dip)}}\sim
 aQ_m\delta b\,
 \frac{\cos\theta}{r^{1+\delta}}.
 }
 \label{eq:app_dipole_anomalous_electric}
\end{equation}
{For $\gamma>0$, this differs from the conventional electric dipole potential, which decays as $r^{-2}$. Its field-strength components exhibit the same anomalous $r^{-2-\delta}$ scaling as the rotation-induced magnetic dipole. At $\gamma=0$, the standard Kerr--Newman electric-dipole behaviour is recovered.}

Finally, in the coupling expansion used in the main text,
\begin{equation}
 V=\frac1r+\gamma v_1+\mathcal O(\gamma^2),\qquad
 \omega=\frac{1-f}{r^2}+\gamma w_1+\mathcal O(\gamma^2),
\end{equation}
the dipole profile becomes
\begin{equation}
 \omega-fV'
=\frac1{r^2}+\gamma\Xi_1+\mathcal O(\gamma^2),
 \qquad \Xi_1=w_1-fv_1'.
\label{eq:app_dipole_perturbative}
\end{equation}
{The same radial functions that determine the rotation-induced magnetic dipole of the electric representative therefore also fix the induced electric dipole of its dyonic duality rotation. Within this family, no additional radial unknown is required.}

\section{Magnetic dipole structure and the departure from Kerr--Newman}
\label{sec:modmax-magnetic-dipole}

The modified magnetic fall-off has a direct interpretation in terms of the
rotation-induced dipole. In Maxwell theory, the moment of a localised
stationary current is
$\boldsymbol{\mu}_{\mathrm{source}}=
\frac12\int\boldsymbol{x}\times\boldsymbol{j}\,d^3x$
in Gaussian units with $c=1$. The same moment determines its far-zone
dipolar field. For a moment aligned with the symmetry axis,
\begin{equation}
 A_\varphi=\frac{\mu}{r}\sin^2\theta,
 \qquad B_{\hat r}=\frac{2\mu}{r^3}\cos\theta,
 \qquad B_{\hat\theta}=\frac{\mu}{r^3}\sin\theta,
 \label{eq:md-maxwell}
\end{equation}
where only the dipole contribution is displayed, and hats denote orthonormal components. Here $A_\varphi$ is the coefficient of the one-form $d\varphi$; the physical component is $A_{\hat\varphi}=A_\varphi/(r\sin\theta)$. For a black hole, the magnetic moment is inferred from the exterior field. Kerr--Newman gives $\mu=Q_ea=Q_eJ/M$.
{A formal charge rescaling would preserve the Maxwell dipole structure in Eq.~\eqref{eq:md-maxwell}, but, as shown below, the resulting Kerr--Newman-type field does not solve the ModMax equations
for nonzero coupling.}

The departure from this structure in ModMax already occurs in flat spacetime. Consider a weak stationary magnetic perturbation of the nonzero electric Coulomb background $\boldsymbol{E}_0=q\hat{\boldsymbol r}/r^2$, with $q=e^{-\gamma}Q_e$. Expanding the ModMax Lagrangian at fixed $\boldsymbol{E}_0$ gives
\begin{equation}
 \begin{aligned}
 \mathcal L&=\frac{e^\gamma}{2}E_0^2
 -\frac{e^{-\gamma}}{2}B_{\hat r}^{\,2}
 -\frac{e^\gamma}{2}B_\perp^2
 +O(B^4/E_0^2),\\
 H_{\hat r}&=e^{-\gamma}B_{\hat r},
 \quad \boldsymbol H_\perp=e^\gamma\boldsymbol B_\perp,
 \end{aligned}
 \label{eq:md-constitutive}
\end{equation}
where $\boldsymbol H=-\partial\mathcal L/\partial\boldsymbol B$
and $\perp$ denotes directions tangent to the spheres. The constitutive
coefficients distinguish radial and tangential magnetic perturbations,
but do not depend on the magnitude of $E_0$. Their difference therefore
persists as $r\to\infty$, reflecting the nonanalyticity of ModMax at
$\mathcal S=\mathcal P=0$. This differs from an analytic nonlinear theory
with a Maxwell weak-field limit, whose constitutive corrections vanish
with the background.

{The presence of the electric background is essential.
For a strictly magnetic configuration, $\boldsymbol E=0$, the ModMax
Lagrangian reduces to
$\mathcal L=-e^{-\gamma}B^2/2$, so pure magnetostatics is
equivalent to Maxwell theory with a rescaled coupling. The anomalous
magnetic response arises instead when magnetic perturbations are
introduced around a nonzero electric field. Although the Coulomb
background tends to zero at infinity, the ratio of the radial and
tangential magnetic response coefficients remains nontrivial.}

Writing $A_\varphi=W(r)\sin^2\theta$, the magnetic field is
\begin{equation}
 B_{\hat r}=\frac{2W}{r^2}\cos\theta,
 \qquad
 B_{\hat\theta}=-\frac{W'}r\sin\theta.
 \label{eq:md-Wfield}
\end{equation}
Outside the source, $\nabla\times\boldsymbol H=0$ then gives
\begin{equation}
 W''-\frac{2e^{-2\gamma}}{r^2}W=0,
 \qquad
 W=Cr^{-\delta},
 \qquad
 \delta=\frac{\sqrt{1+8e^{-2\gamma}}-1}{2},
 \label{eq:md-flat-mode}
\end{equation}
where the growing solution has been excluded. This exponent is the
$\ell=1$ member of the flat-space axial multipole exponents obtained in
Ref.~{\cite{BeltranJimenezBettoniBrax:2023}}.
{More generally, the decaying axial multipoles are
characterised by
$\delta_\ell(\delta_\ell+1)=e^{-2\gamma}\ell(\ell+1)$,
which reduces to the usual integer multipole hierarchy
$\delta_\ell=\ell$ in the Maxwell limit.}
For finite $\gamma>0$,
$0<\delta<1$, and hence
\begin{equation}
 B_{\hat r}=\frac{2C\cos\theta}{r^{2+\delta}},
 \qquad
 B_{\hat\theta}=\frac{\delta C\sin\theta}{r^{2+\delta}}.
 \label{eq:md-anomalous-field}
\end{equation}
The field remains an angular dipole, but both its radial decay and the
relative coefficients of its components differ from Maxwell's.
Moreover, $|\boldsymbol B|/|\boldsymbol E_0|=O(r^{-\delta})$,
so the perturbation remains controlled at large radius.

The distinction matters for interpreting the magnetic moment.
Applying the Maxwell extraction separately to the two components yields
\begin{equation}
 \mu_{\mathrm{eff}}^{(r)}(r)
 =\frac{r^3B_{\hat r}}{2\cos\theta}=Cr^{1-\delta},
 \qquad
 \mu_{\mathrm{eff}}^{(\theta)}(r)
 =\frac{r^3B_{\hat\theta}}{\sin\theta}
 =\delta Cr^{1-\delta}.
 \label{eq:md-effective-moments}
\end{equation}
For a nonzero mode, these neither coincide nor approach a finite constant
when $\gamma>0$. Thus the anomalous tail cannot be captured by assigning
a different constant dipole moment or gyromagnetic ratio to a Maxwell
profile. A prescribed localised current still has a well-defined source
moment; its identification with a conventional $1/r$ coefficient of
$A_\varphi$ is what fails. The amplitude $C$ instead characterises the
modified dipolar mode.

{Unlike the conventional Maxwell magnetic moment,
this amplitude carries a coupling-dependent radial scaling. It
reduces to the usual dipole moment when $\gamma=0$, but for
$\gamma>0$ it should be regarded as the coefficient of an anomalous
dipolar mode rather than as a constant Maxwell-like magnetic moment.}

For the rotating black hole, $W=aqV$ and the equations at first order in
$a$, without expanding in $\gamma$, are
\begin{equation}
 (fV')'-\frac{2e^{-2\gamma}}{r^2}V=\omega',
 \qquad r^4\omega'=4KV-6M,
 \label{eq:md-bh-system}
\end{equation}
with $f=1-2M/r+K/r^2$ and $K=e^{-\gamma}Q_e^2$.
Their leading asymptotic homogeneous operator is precisely the flat-space
operator in Eq.~\eqref{eq:md-flat-mode}. The regular nonextremal solution
therefore has the same exponent, $V\sim br^{-\delta}$, with the nonzero
amplitude $b$ fixed by horizon regularity and the absence of the growing
mode. Gravity changes the radial profile and selects the amplitude,
while the exponent is inherited from the flat-space constitutive response.

{In terms of the horizon connection coefficient
introduced in \autoref{sec:finite}, the asymptotic magnetic
amplitude of the rotating black hole is
$C=aqb=aqC_\gamma r_+^{\delta-1}$.
Thus, the exponent is determined entirely by the ModMax constitutive
law, whereas the amplitude depends on the black hole parameters and
is fixed by the boundary-value problem. In the Maxwell limit,
$C_\gamma\to1$ and $\delta\to1$, recovering the Kerr--Newman
magnetic moment $C=Q_ea$.}

This also provides an asymptotic explanation of the obstruction to the
restricted Kerr--Newman ansatz discussed in
Ref.~{\cite{Kubiznak:2022vft}}.
A nonzero Maxwell tail $V=v_0/r+\cdots$
produces $2(1-e^{-2\gamma})v_0/r^3$ in the electromagnetic equation,
whereas $\omega'$ starts at order $r^{-4}$. Charge screening cannot remove
this mismatch. More strongly, imposing
$\omega_{\mathrm{KN}}=(1-f)/r^2$ forces $V_{\mathrm{KN}}=1/r$
through the Einstein equation for $K\ne0$, leaving the exact residual
\begin{equation}
 (fV_{\mathrm{KN}}')'
 -\frac{2e^{-2\gamma}}{r^2}V_{\mathrm{KN}}
 -\omega_{\mathrm{KN}}'
 =\frac{2(1-e^{-2\gamma})}{r^3}.
 \label{eq:md-KN-residual}
\end{equation}
The independent profiles $V$ and $\omega$ in our ansatz supply the freedom
needed to accommodate this magnetic structure; the boundary conditions
subsequently fix them. Their asymptotics preserve
$g_{t\varphi}=-2J\sin^2\theta/r+o(r^{-1})$. The obstruction concerns
ans\"atze that enforce the Kerr–Newman magnetic profile or its linked
frame-dragging relation, and it already applies at first order to any smooth
rotating family connected to the charged static seed.

{It does not constitute a no-go theorem for rotating
Einstein--ModMax black holes. Rather, it identifies the non-Maxwell
asymptotic magnetic structure that any such solution must accommodate
on the branch continuously connected to the static charged background.}

\bibliography{biblio}

\begin{thebibliography}{31}%
\makeatletter
\providecommand \@ifxundefined [1]{%
 \@ifx{#1\undefined}
}%
\providecommand \@ifnum [1]{%
 \ifnum #1\expandafter \@firstoftwo
 \else \expandafter \@secondoftwo
 \fi
}%
\providecommand \@ifx [1]{%
 \ifx #1\expandafter \@firstoftwo
 \else \expandafter \@secondoftwo
 \fi
}%
\providecommand \natexlab [1]{#1}%
\providecommand \enquote  [1]{``#1''}%
\providecommand \bibnamefont  [1]{#1}%
\providecommand \bibfnamefont [1]{#1}%
\providecommand \citenamefont [1]{#1}%
\providecommand \href@noop [0]{\@secondoftwo}%
\providecommand \href [0]{\begingroup \@sanitize@url \@href}%
\providecommand \@href[1]{\@@startlink{#1}\@@href}%
\providecommand \@@href[1]{\endgroup#1\@@endlink}%
\providecommand \@sanitize@url [0]{\catcode `\\12\catcode `\$12\catcode `\&12\catcode `\#12\catcode `\^12\catcode `\_12\catcode `\%12\relax}%
\providecommand \@@startlink[1]{}%
\providecommand \@@endlink[0]{}%
\providecommand \url  [0]{\begingroup\@sanitize@url \@url }%
\providecommand \@url [1]{\endgroup\@href {#1}{\urlprefix }}%
\providecommand \urlprefix  [0]{URL }%
\providecommand \Eprint [0]{\href }%
\providecommand \doibase [0]{https://doi.org/}%
\providecommand \selectlanguage [0]{\@gobble}%
\providecommand \bibinfo  [0]{\@secondoftwo}%
\providecommand \bibfield  [0]{\@secondoftwo}%
\providecommand \translation [1]{[#1]}%
\providecommand \BibitemOpen [0]{}%
\providecommand \bibitemStop [0]{}%
\providecommand \bibitemNoStop [0]{.\EOS\space}%
\providecommand \EOS [0]{\spacefactor3000\relax}%
\providecommand \BibitemShut  [1]{\csname bibitem#1\endcsname}%
\let\auto@bib@innerbib\@empty
\bibitem [{\citenamefont {Born}\ and\ \citenamefont {Infeld}(1934)}]{Born:1934gh}%
  \BibitemOpen
  \bibfield  {author} {\bibinfo {author} {\bibfnamefont {M.}~\bibnamefont {Born}}\ and\ \bibinfo {author} {\bibfnamefont {L.}~\bibnamefont {Infeld}},\ }\href {https://doi.org/10.1098/rspa.1934.0059} {\bibfield  {journal} {\bibinfo  {journal} {Proc. Roy. Soc. Lond. A}\ }\textbf {\bibinfo {volume} {144}},\ \bibinfo {pages} {425} (\bibinfo {year} {1934})}\BibitemShut {NoStop}%
\bibitem [{\citenamefont {Heisenberg}\ and\ \citenamefont {Euler}(1936)}]{Heisenberg:1936nmg}%
  \BibitemOpen
  \bibfield  {author} {\bibinfo {author} {\bibfnamefont {W.}~\bibnamefont {Heisenberg}}\ and\ \bibinfo {author} {\bibfnamefont {H.}~\bibnamefont {Euler}},\ }\href {https://doi.org/10.1007/BF01343663} {\bibfield  {journal} {\bibinfo  {journal} {Z. Phys.}\ }\textbf {\bibinfo {volume} {98}},\ \bibinfo {pages} {714} (\bibinfo {year} {1936})},\ \Eprint {https://arxiv.org/abs/physics/0605038} {arXiv:physics/0605038} \BibitemShut {NoStop}%
\bibitem [{\citenamefont {Bandos}\ \emph {et~al.}(2020)\citenamefont {Bandos}, \citenamefont {Lechner}, \citenamefont {Sorokin},\ and\ \citenamefont {Townsend}}]{Bandos:2020jsw}%
  \BibitemOpen
  \bibfield  {author} {\bibinfo {author} {\bibfnamefont {I.}~\bibnamefont {Bandos}}, \bibinfo {author} {\bibfnamefont {K.}~\bibnamefont {Lechner}}, \bibinfo {author} {\bibfnamefont {D.}~\bibnamefont {Sorokin}},\ and\ \bibinfo {author} {\bibfnamefont {P.~K.}\ \bibnamefont {Townsend}},\ }\href {https://doi.org/10.1103/PhysRevD.102.121703} {\bibfield  {journal} {\bibinfo  {journal} {Phys. Rev. D}\ }\textbf {\bibinfo {volume} {102}},\ \bibinfo {pages} {121703} (\bibinfo {year} {2020})},\ \Eprint {https://arxiv.org/abs/2007.09092} {arXiv:2007.09092 [hep-th]} \BibitemShut {NoStop}%
\bibitem [{\citenamefont {Bandos}\ \emph {et~al.}(2021{\natexlab{a}})\citenamefont {Bandos}, \citenamefont {Lechner}, \citenamefont {Sorokin},\ and\ \citenamefont {Townsend}}]{Bandos:2021rqy}%
  \BibitemOpen
  \bibfield  {author} {\bibinfo {author} {\bibfnamefont {I.}~\bibnamefont {Bandos}}, \bibinfo {author} {\bibfnamefont {K.}~\bibnamefont {Lechner}}, \bibinfo {author} {\bibfnamefont {D.}~\bibnamefont {Sorokin}},\ and\ \bibinfo {author} {\bibfnamefont {P.~K.}\ \bibnamefont {Townsend}},\ }\href {https://doi.org/10.1007/JHEP10(2021)031} {\bibfield  {journal} {\bibinfo  {journal} {JHEP}\ }\textbf {\bibinfo {volume} {2021}}\bibfield  {number} {\bibinfo  {number} { (10)},\ \bibinfo {pages} {031}},\ }\Eprint {https://arxiv.org/abs/2106.07547} {arXiv:2106.07547 [hep-th]} \BibitemShut {NoStop}%
\bibitem [{\citenamefont {Bandos}\ \emph {et~al.}(2021{\natexlab{b}})\citenamefont {Bandos}, \citenamefont {Lechner}, \citenamefont {Sorokin},\ and\ \citenamefont {Townsend}}]{Bandos:2020hgy}%
  \BibitemOpen
  \bibfield  {author} {\bibinfo {author} {\bibfnamefont {I.}~\bibnamefont {Bandos}}, \bibinfo {author} {\bibfnamefont {K.}~\bibnamefont {Lechner}}, \bibinfo {author} {\bibfnamefont {D.}~\bibnamefont {Sorokin}},\ and\ \bibinfo {author} {\bibfnamefont {P.~K.}\ \bibnamefont {Townsend}},\ }\href {https://doi.org/10.1007/JHEP03(2021)022} {\bibfield  {journal} {\bibinfo  {journal} {JHEP}\ }\textbf {\bibinfo {volume} {2021}}\bibfield  {number} {\bibinfo  {number} { (3)},\ \bibinfo {pages} {022}},\ }\Eprint {https://arxiv.org/abs/2012.09286} {arXiv:2012.09286 [hep-th]} \BibitemShut {NoStop}%
\bibitem [{\citenamefont {Ferko}\ \emph {et~al.}(2022{\natexlab{a}})\citenamefont {Ferko}, \citenamefont {Smith},\ and\ \citenamefont {Tartaglino-Mazzucchelli}}]{Ferko:2022iru}%
  \BibitemOpen
  \bibfield  {author} {\bibinfo {author} {\bibfnamefont {C.}~\bibnamefont {Ferko}}, \bibinfo {author} {\bibfnamefont {L.}~\bibnamefont {Smith}},\ and\ \bibinfo {author} {\bibfnamefont {G.}~\bibnamefont {Tartaglino-Mazzucchelli}},\ }\href {https://doi.org/10.21468/SciPostPhys.13.2.012} {\bibfield  {journal} {\bibinfo  {journal} {SciPost Phys.}\ }\textbf {\bibinfo {volume} {13}},\ \bibinfo {pages} {012} (\bibinfo {year} {2022}{\natexlab{a}})},\ \Eprint {https://arxiv.org/abs/2203.01085} {arXiv:2203.01085 [hep-th]} \BibitemShut {NoStop}%
\bibitem [{\citenamefont {Ferko}\ \emph {et~al.}(2022{\natexlab{b}})\citenamefont {Ferko}, \citenamefont {Sfondrini}, \citenamefont {Smith},\ and\ \citenamefont {Tartaglino-Mazzucchelli}}]{Ferko:2022cix}%
  \BibitemOpen
  \bibfield  {author} {\bibinfo {author} {\bibfnamefont {C.}~\bibnamefont {Ferko}}, \bibinfo {author} {\bibfnamefont {A.}~\bibnamefont {Sfondrini}}, \bibinfo {author} {\bibfnamefont {L.}~\bibnamefont {Smith}},\ and\ \bibinfo {author} {\bibfnamefont {G.}~\bibnamefont {Tartaglino-Mazzucchelli}},\ }\href {https://doi.org/10.1103/PhysRevLett.129.201604} {\bibfield  {journal} {\bibinfo  {journal} {Phys. Rev. Lett.}\ }\textbf {\bibinfo {volume} {129}},\ \bibinfo {pages} {201604} (\bibinfo {year} {2022}{\natexlab{b}})},\ \Eprint {https://arxiv.org/abs/2206.10515} {arXiv:2206.10515 [hep-th]} \BibitemShut {NoStop}%
\bibitem [{\citenamefont {Dassy}\ and\ \citenamefont {Govaerts}(2021)}]{Dassy:2021ulu}%
  \BibitemOpen
  \bibfield  {author} {\bibinfo {author} {\bibfnamefont {C.}~\bibnamefont {Dassy}}\ and\ \bibinfo {author} {\bibfnamefont {J.}~\bibnamefont {Govaerts}},\ }\href {https://doi.org/10.1088/1751-8121/abfffe} {\bibfield  {journal} {\bibinfo  {journal} {J. Phys. A}\ }\textbf {\bibinfo {volume} {54}},\ \bibinfo {pages} {265201} (\bibinfo {year} {2021})},\ \Eprint {https://arxiv.org/abs/2105.05802} {arXiv:2105.05802 [physics.gen-ph]} \BibitemShut {NoStop}%
\bibitem [{\citenamefont {Lechner}\ \emph {et~al.}(2022)\citenamefont {Lechner}, \citenamefont {Marchetti}, \citenamefont {Sainaghi},\ and\ \citenamefont {Sorokin}}]{Lechner:2022qhb}%
  \BibitemOpen
  \bibfield  {author} {\bibinfo {author} {\bibfnamefont {K.}~\bibnamefont {Lechner}}, \bibinfo {author} {\bibfnamefont {P.}~\bibnamefont {Marchetti}}, \bibinfo {author} {\bibfnamefont {A.}~\bibnamefont {Sainaghi}},\ and\ \bibinfo {author} {\bibfnamefont {D.~P.}\ \bibnamefont {Sorokin}},\ }\href {https://doi.org/10.1103/PhysRevD.106.016009} {\bibfield  {journal} {\bibinfo  {journal} {Phys. Rev. D}\ }\textbf {\bibinfo {volume} {106}},\ \bibinfo {pages} {016009} (\bibinfo {year} {2022})},\ \Eprint {https://arxiv.org/abs/2206.04657} {arXiv:2206.04657 [hep-th]} \BibitemShut {NoStop}%
\bibitem [{\citenamefont {Neves}\ \emph {et~al.}(2023)\citenamefont {Neves}, \citenamefont {Gaete}, \citenamefont {Ospedal},\ and\ \citenamefont {Helay{\"e}l-Neto}}]{Neves:2022jqq}%
  \BibitemOpen
  \bibfield  {author} {\bibinfo {author} {\bibfnamefont {M.~J.}\ \bibnamefont {Neves}}, \bibinfo {author} {\bibfnamefont {P.}~\bibnamefont {Gaete}}, \bibinfo {author} {\bibfnamefont {L.~P.~R.}\ \bibnamefont {Ospedal}},\ and\ \bibinfo {author} {\bibfnamefont {J.~A.}\ \bibnamefont {Helay{\"e}l-Neto}},\ }\href {https://doi.org/10.1103/PhysRevD.107.075019} {\bibfield  {journal} {\bibinfo  {journal} {Phys. Rev. D}\ }\textbf {\bibinfo {volume} {107}},\ \bibinfo {pages} {075019} (\bibinfo {year} {2023})},\ \Eprint {https://arxiv.org/abs/2209.09361} {arXiv:2209.09361 [hep-th]} \BibitemShut {NoStop}%
\bibitem [{\citenamefont {Kuzenko}\ and\ \citenamefont {Raptakis}(2024)}]{Kuzenko:2024zra}%
  \BibitemOpen
  \bibfield  {author} {\bibinfo {author} {\bibfnamefont {S.~M.}\ \bibnamefont {Kuzenko}}\ and\ \bibinfo {author} {\bibfnamefont {E.~S.~N.}\ \bibnamefont {Raptakis}},\ }\href {https://doi.org/10.1007/JHEP06(2024)162} {\bibfield  {journal} {\bibinfo  {journal} {JHEP}\ }\textbf {\bibinfo {volume} {2024}}\bibfield  {number} {\bibinfo  {number} { (6)},\ \bibinfo {pages} {162}},\ }\Eprint {https://arxiv.org/abs/2404.09108} {arXiv:2404.09108 [hep-th]} \BibitemShut {NoStop}%
\bibitem [{\citenamefont {Sorge}(2024)}]{Sorge:2024gag}%
  \BibitemOpen
  \bibfield  {author} {\bibinfo {author} {\bibfnamefont {F.}~\bibnamefont {Sorge}},\ }\href {https://doi.org/10.1103/PhysRevD.110.116006} {\bibfield  {journal} {\bibinfo  {journal} {Phys. Rev. D}\ }\textbf {\bibinfo {volume} {110}},\ \bibinfo {pages} {116006} (\bibinfo {year} {2024})}\BibitemShut {NoStop}%
\bibitem [{\citenamefont {Buchbinder}\ \emph {et~al.}(2026)\citenamefont {Buchbinder}, \citenamefont {Grasso},\ and\ \citenamefont {Pinelli}}]{Buchbinder:2026nop}%
  \BibitemOpen
  \bibfield  {author} {\bibinfo {author} {\bibfnamefont {E.~I.}\ \bibnamefont {Buchbinder}}, \bibinfo {author} {\bibfnamefont {D.~T.}\ \bibnamefont {Grasso}},\ and\ \bibinfo {author} {\bibfnamefont {J.~R.}\ \bibnamefont {Pinelli}},\ }\href {https://doi.org/10.1007/JHEP05(2026)017} {\bibfield  {journal} {\bibinfo  {journal} {JHEP}\ }\textbf {\bibinfo {volume} {2026}}\bibfield  {number} {\bibinfo  {number} { (5)},\ \bibinfo {pages} {017}},\ }\Eprint {https://arxiv.org/abs/2601.19339} {arXiv:2601.19339 [hep-th]} \BibitemShut {NoStop}%
\bibitem [{\citenamefont {Barrientos}\ \emph {et~al.}(2025{\natexlab{a}})\citenamefont {Barrientos}, \citenamefont {C{\'a}ceres}, \citenamefont {Diaz},\ and\ \citenamefont {Hernandez-Vera}}]{Barrientos:2025rde}%
  \BibitemOpen
  \bibfield  {author} {\bibinfo {author} {\bibfnamefont {J.}~\bibnamefont {Barrientos}}, \bibinfo {author} {\bibfnamefont {N.}~\bibnamefont {C{\'a}ceres}}, \bibinfo {author} {\bibfnamefont {F.}~\bibnamefont {Diaz}},\ and\ \bibinfo {author} {\bibfnamefont {U.}~\bibnamefont {Hernandez-Vera}},\ }\href {https://doi.org/10.1103/d3tj-4nd8} {\bibfield  {journal} {\bibinfo  {journal} {Phys. Rev. D}\ }\textbf {\bibinfo {volume} {112}},\ \bibinfo {pages} {086018} (\bibinfo {year} {2025}{\natexlab{a}})},\ \Eprint {https://arxiv.org/abs/2506.02884} {arXiv:2506.02884 [hep-th]} \BibitemShut {NoStop}%
\bibitem [{\citenamefont {Colip{\'i}-Marchant}\ \emph {et~al.}(2023)\citenamefont {Colip{\'i}-Marchant}, \citenamefont {Corral}, \citenamefont {Flores-Alfonso},\ and\ \citenamefont {Sanhueza}}]{ColipiMarchant:2023AxialAnomaly}%
  \BibitemOpen
  \bibfield  {author} {\bibinfo {author} {\bibfnamefont {F.}~\bibnamefont {Colip{\'i}-Marchant}}, \bibinfo {author} {\bibfnamefont {C.}~\bibnamefont {Corral}}, \bibinfo {author} {\bibfnamefont {D.}~\bibnamefont {Flores-Alfonso}},\ and\ \bibinfo {author} {\bibfnamefont {L.}~\bibnamefont {Sanhueza}},\ }\href {https://doi.org/10.1103/PhysRevD.107.104042} {\bibfield  {journal} {\bibinfo  {journal} {Phys. Rev. D}\ }\textbf {\bibinfo {volume} {107}},\ \bibinfo {pages} {104042} (\bibinfo {year} {2023})},\ \Eprint {https://arxiv.org/abs/2302.09162} {arXiv:2302.09162 [hep-th]} \BibitemShut {NoStop}%
\bibitem [{\citenamefont {Flores-Alfonso}\ \emph {et~al.}(2021{\natexlab{a}})\citenamefont {Flores-Alfonso}, \citenamefont {Gonz{\'a}lez-Morales}, \citenamefont {Linares},\ and\ \citenamefont {Maceda}}]{Flores-Alfonso:2020euz}%
  \BibitemOpen
  \bibfield  {author} {\bibinfo {author} {\bibfnamefont {D.}~\bibnamefont {Flores-Alfonso}}, \bibinfo {author} {\bibfnamefont {B.~A.}\ \bibnamefont {Gonz{\'a}lez-Morales}}, \bibinfo {author} {\bibfnamefont {R.}~\bibnamefont {Linares}},\ and\ \bibinfo {author} {\bibfnamefont {M.}~\bibnamefont {Maceda}},\ }\href {https://doi.org/10.1016/j.physletb.2020.136011} {\bibfield  {journal} {\bibinfo  {journal} {Phys. Lett. B}\ }\textbf {\bibinfo {volume} {812}},\ \bibinfo {pages} {136011} (\bibinfo {year} {2021}{\natexlab{a}})},\ \Eprint {https://arxiv.org/abs/2011.10836} {arXiv:2011.10836 [gr-qc]} \BibitemShut {NoStop}%
\bibitem [{\citenamefont {Ballon~Bordo}\ \emph {et~al.}(2021)\citenamefont {Ballon~Bordo}, \citenamefont {Kubiz{\v{n}}{\'a}k},\ and\ \citenamefont {Perche}}]{BallonBordo:2020jtw}%
  \BibitemOpen
  \bibfield  {author} {\bibinfo {author} {\bibfnamefont {A.}~\bibnamefont {Ballon~Bordo}}, \bibinfo {author} {\bibfnamefont {D.}~\bibnamefont {Kubiz{\v{n}}{\'a}k}},\ and\ \bibinfo {author} {\bibfnamefont {T.~R.}\ \bibnamefont {Perche}},\ }\href {https://doi.org/10.1016/j.physletb.2021.136312} {\bibfield  {journal} {\bibinfo  {journal} {Phys. Lett. B}\ }\textbf {\bibinfo {volume} {817}},\ \bibinfo {pages} {136312} (\bibinfo {year} {2021})},\ \Eprint {https://arxiv.org/abs/2011.13398} {arXiv:2011.13398 [hep-th]} \BibitemShut {NoStop}%
\bibitem [{\citenamefont {Flores-Alfonso}\ \emph {et~al.}(2021{\natexlab{b}})\citenamefont {Flores-Alfonso}, \citenamefont {Linares},\ and\ \citenamefont {Maceda}}]{Flores-Alfonso:2020nnd}%
  \BibitemOpen
  \bibfield  {author} {\bibinfo {author} {\bibfnamefont {D.}~\bibnamefont {Flores-Alfonso}}, \bibinfo {author} {\bibfnamefont {R.}~\bibnamefont {Linares}},\ and\ \bibinfo {author} {\bibfnamefont {M.}~\bibnamefont {Maceda}},\ }\href {https://doi.org/10.1007/JHEP09(2021)104} {\bibfield  {journal} {\bibinfo  {journal} {JHEP}\ }\textbf {\bibinfo {volume} {2021}}\bibfield  {number} {\bibinfo  {number} { (9)},\ \bibinfo {pages} {104}},\ }\Eprint {https://arxiv.org/abs/2012.03416} {arXiv:2012.03416 [gr-qc]} \BibitemShut {NoStop}%
\bibitem [{\citenamefont {Barrientos}\ \emph {et~al.}(2022)\citenamefont {Barrientos}, \citenamefont {Cisterna}, \citenamefont {Kubiznak},\ and\ \citenamefont {Oliva}}]{Barrientos:2022bzm}%
  \BibitemOpen
  \bibfield  {author} {\bibinfo {author} {\bibfnamefont {J.}~\bibnamefont {Barrientos}}, \bibinfo {author} {\bibfnamefont {A.}~\bibnamefont {Cisterna}}, \bibinfo {author} {\bibfnamefont {D.}~\bibnamefont {Kubiznak}},\ and\ \bibinfo {author} {\bibfnamefont {J.}~\bibnamefont {Oliva}},\ }\href {https://doi.org/10.1016/j.physletb.2022.137447} {\bibfield  {journal} {\bibinfo  {journal} {Phys. Lett. B}\ }\textbf {\bibinfo {volume} {834}},\ \bibinfo {pages} {137447} (\bibinfo {year} {2022})},\ \Eprint {https://arxiv.org/abs/2205.15777} {arXiv:2205.15777 [gr-qc]} \BibitemShut {NoStop}%
\bibitem [{\citenamefont {Barrientos}\ \emph {et~al.}(2025{\natexlab{b}})\citenamefont {Barrientos}, \citenamefont {Cisterna}, \citenamefont {Hassaine},\ and\ \citenamefont {Pallikaris}}]{Barrientos:2024umq}%
  \BibitemOpen
  \bibfield  {author} {\bibinfo {author} {\bibfnamefont {J.}~\bibnamefont {Barrientos}}, \bibinfo {author} {\bibfnamefont {A.}~\bibnamefont {Cisterna}}, \bibinfo {author} {\bibfnamefont {M.}~\bibnamefont {Hassaine}},\ and\ \bibinfo {author} {\bibfnamefont {K.}~\bibnamefont {Pallikaris}},\ }\href {https://doi.org/10.1016/j.physletb.2024.139214} {\bibfield  {journal} {\bibinfo  {journal} {Phys. Lett. B}\ }\textbf {\bibinfo {volume} {860}},\ \bibinfo {pages} {139214} (\bibinfo {year} {2025}{\natexlab{b}})},\ \Eprint {https://arxiv.org/abs/2409.12336} {arXiv:2409.12336 [gr-qc]} \BibitemShut {NoStop}%
\bibitem [{\citenamefont {Bokuli{\'c}}\ and\ \citenamefont {Herdeiro}(2025{\natexlab{a}})}]{Bokulic:2025usc}%
  \BibitemOpen
  \bibfield  {author} {\bibinfo {author} {\bibfnamefont {A.}~\bibnamefont {Bokuli{\'c}}}\ and\ \bibinfo {author} {\bibfnamefont {C.~A.~R.}\ \bibnamefont {Herdeiro}},\ }\href {https://doi.org/10.1103/PhysRevD.111.064046} {\bibfield  {journal} {\bibinfo  {journal} {Phys. Rev. D}\ }\textbf {\bibinfo {volume} {111}},\ \bibinfo {pages} {064046} (\bibinfo {year} {2025}{\natexlab{a}})},\ \Eprint {https://arxiv.org/abs/2501.04779} {arXiv:2501.04779 [gr-qc]} \BibitemShut {NoStop}%
\bibitem [{\citenamefont {Bokuli{\'c}}\ and\ \citenamefont {Herdeiro}(2025{\natexlab{b}})}]{Bokulic:2025ucy}%
  \BibitemOpen
  \bibfield  {author} {\bibinfo {author} {\bibfnamefont {A.}~\bibnamefont {Bokuli{\'c}}}\ and\ \bibinfo {author} {\bibfnamefont {C.~A.~R.}\ \bibnamefont {Herdeiro}},\ }\href {https://doi.org/10.1007/JHEP10(2025)091} {\bibfield  {journal} {\bibinfo  {journal} {JHEP}\ }\textbf {\bibinfo {volume} {2025}}\bibfield  {number} {\bibinfo  {number} { (10)},\ \bibinfo {pages} {091}},\ }\Eprint {https://arxiv.org/abs/2507.16926} {arXiv:2507.16926 [gr-qc]} \BibitemShut {NoStop}%
\bibitem [{\citenamefont {Bixano}\ and\ \citenamefont {Matos}(2026{\natexlab{a}})}]{Bixano:2026ouq}%
  \BibitemOpen
  \bibfield  {author} {\bibinfo {author} {\bibfnamefont {L.}~\bibnamefont {Bixano}}\ and\ \bibinfo {author} {\bibfnamefont {T.}~\bibnamefont {Matos}},\ }\href@noop {} {\bibinfo {title} {{Generalized Einstein-ModMax-ScalarField theories and new exact solutions}}} (\bibinfo {year} {2026}{\natexlab{a}}),\ \Eprint {https://arxiv.org/abs/2603.26073} {arXiv:2603.26073 [gr-qc]} \BibitemShut {NoStop}%
\bibitem [{\citenamefont {Bixano}\ and\ \citenamefont {Matos}(2026{\natexlab{b}})}]{Bixano:2026duf}%
  \BibitemOpen
  \bibfield  {author} {\bibinfo {author} {\bibfnamefont {L.}~\bibnamefont {Bixano}}\ and\ \bibinfo {author} {\bibfnamefont {T.}~\bibnamefont {Matos}},\ }\href {https://doi.org/10.1088/1361-6382/aeb8dc} {\bibfield  {journal} {\bibinfo  {journal} {Class. Quant. Grav.}\ }\textbf {\bibinfo {volume} {43}},\ \bibinfo {pages} {215006} (\bibinfo {year} {2026}{\natexlab{b}})},\ \Eprint {https://arxiv.org/abs/2605.17843} {arXiv:2605.17843 [gr-qc]} \BibitemShut {NoStop}%
\bibitem [{\citenamefont {D{\'\i}az~Garc{\'\i}a}(2022)}]{DiazGarcia:2022jpc}%
  \BibitemOpen
  \bibfield  {author} {\bibinfo {author} {\bibfnamefont {A.~A.}\ \bibnamefont {D{\'\i}az~Garc{\'\i}a}},\ }\href {https://doi.org/10.1016/j.aop.2022.168880} {\bibfield  {journal} {\bibinfo  {journal} {Annals Phys.}\ }\textbf {\bibinfo {volume} {441}},\ \bibinfo {pages} {168880} (\bibinfo {year} {2022})},\ \Eprint {https://arxiv.org/abs/2201.10682} {arXiv:2201.10682 [gr-qc]} \BibitemShut {NoStop}%
\bibitem [{\citenamefont {Ay{\'o}n-Beato}(2024)}]{Ayon-Beato:2022dwg}%
  \BibitemOpen
  \bibfield  {author} {\bibinfo {author} {\bibfnamefont {E.}~\bibnamefont {Ay{\'o}n-Beato}},\ }\href {https://doi.org/10.1016/j.aop.2024.169771} {\bibfield  {journal} {\bibinfo  {journal} {Annals Phys.}\ }\textbf {\bibinfo {volume} {469}},\ \bibinfo {pages} {169771} (\bibinfo {year} {2024})},\ \Eprint {https://arxiv.org/abs/2203.12809} {arXiv:2203.12809 [gr-qc]} \BibitemShut {NoStop}%
\bibitem [{\citenamefont {Kubiznak}\ \emph {et~al.}(2022)\citenamefont {Kubiznak}, \citenamefont {Tahamtan},\ and\ \citenamefont {Svitek}}]{Kubiznak:2022vft}%
  \BibitemOpen
  \bibfield  {author} {\bibinfo {author} {\bibfnamefont {D.}~\bibnamefont {Kubiznak}}, \bibinfo {author} {\bibfnamefont {T.}~\bibnamefont {Tahamtan}},\ and\ \bibinfo {author} {\bibfnamefont {O.}~\bibnamefont {Svitek}},\ }\href {https://doi.org/10.1103/PhysRevD.105.104064} {\bibfield  {journal} {\bibinfo  {journal} {Phys. Rev. D}\ }\textbf {\bibinfo {volume} {105}},\ \bibinfo {pages} {104064} (\bibinfo {year} {2022})},\ \Eprint {https://arxiv.org/abs/2203.01919} {arXiv:2203.01919 [gr-qc]} \BibitemShut {NoStop}%
\bibitem [{\citenamefont {Hale}\ \emph {et~al.}(2023)\citenamefont {Hale}, \citenamefont {Kubiz{\v{n}}{\'a}k}, \citenamefont {Sv{\'\i}tek},\ and\ \citenamefont {Tahamtan}}]{Hale:2023dpf}%
  \BibitemOpen
  \bibfield  {author} {\bibinfo {author} {\bibfnamefont {T.}~\bibnamefont {Hale}}, \bibinfo {author} {\bibfnamefont {D.}~\bibnamefont {Kubiz{\v{n}}{\'a}k}}, \bibinfo {author} {\bibfnamefont {O.}~\bibnamefont {Sv{\'\i}tek}},\ and\ \bibinfo {author} {\bibfnamefont {T.}~\bibnamefont {Tahamtan}},\ }\href {https://doi.org/10.1103/PhysRevD.107.124031} {\bibfield  {journal} {\bibinfo  {journal} {Phys. Rev. D}\ }\textbf {\bibinfo {volume} {107}},\ \bibinfo {pages} {124031} (\bibinfo {year} {2023})},\ \Eprint {https://arxiv.org/abs/2303.16928} {arXiv:2303.16928 [gr-qc]} \BibitemShut {NoStop}%
\bibitem [{\citenamefont {Hale}\ \emph {et~al.}(2026)\citenamefont {Hale}, \citenamefont {Hennigar},\ and\ \citenamefont {Kubiz{\v{n}}{\'a}k}}]{Hale:2025urg}%
  \BibitemOpen
  \bibfield  {author} {\bibinfo {author} {\bibfnamefont {T.}~\bibnamefont {Hale}}, \bibinfo {author} {\bibfnamefont {R.~A.}\ \bibnamefont {Hennigar}},\ and\ \bibinfo {author} {\bibfnamefont {D.}~\bibnamefont {Kubiz{\v{n}}{\'a}k}},\ }\href {https://doi.org/10.1007/JHEP01(2026)155} {\bibfield  {journal} {\bibinfo  {journal} {JHEP}\ }\textbf {\bibinfo {volume} {2026}}\bibfield  {number} {\bibinfo  {number} { (1)},\ \bibinfo {pages} {155}},\ }\Eprint {https://arxiv.org/abs/2509.13099} {arXiv:2509.13099 [gr-qc]} \BibitemShut {NoStop}%
\bibitem [{\citenamefont {Beltran~Jimenez}\ \emph {et~al.}(2023)\citenamefont {Beltran~Jimenez}, \citenamefont {Bettoni},\ and\ \citenamefont {Brax}}]{BeltranJimenezBettoniBrax:2023}%
  \BibitemOpen
  \bibfield  {author} {\bibinfo {author} {\bibfnamefont {J.}~\bibnamefont {Beltran~Jimenez}}, \bibinfo {author} {\bibfnamefont {D.}~\bibnamefont {Bettoni}},\ and\ \bibinfo {author} {\bibfnamefont {P.}~\bibnamefont {Brax}},\ }\href {https://doi.org/10.1007/JHEP02(2023)009} {\bibfield  {journal} {\bibinfo  {journal} {JHEP}\ }\textbf {\bibinfo {volume} {2023}}\bibfield  {number} {\bibinfo  {number} { (2)},\ \bibinfo {pages} {009}},\ }\Eprint {https://arxiv.org/abs/2211.16404} {arXiv:2211.16404 [hep-th]} \BibitemShut {NoStop}%
\bibitem [{\citenamefont {Minguzzi}(2010)}]{Minguzzi:2010juk}%
  \BibitemOpen
  \bibfield  {author} {\bibinfo {author} {\bibfnamefont {E.}~\bibnamefont {Minguzzi}},\ }\href {https://doi.org/10.1007/s00220-010-1048-1} {\bibfield  {journal} {\bibinfo  {journal} {Commun. Math. Phys.}\ }\textbf {\bibinfo {volume} {298}},\ \bibinfo {pages} {855} (\bibinfo {year} {2010})},\ \Eprint {https://arxiv.org/abs/0909.0890} {arXiv:0909.0890 [gr-qc]} \BibitemShut {NoStop}%
\end{thebibliography}%

\end{document}